\documentclass[5p]{elsarticle}
\usepackage{subfigure}
\usepackage{amssymb}
\usepackage{amsfonts}
\usepackage{amsmath}
\usepackage{amsthm}
\usepackage{epsfig}
\usepackage{graphicx}
\usepackage[usenames,dvipsnames]{color}
\usepackage[latin1]{inputenc}
\usepackage{booktabs}
\usepackage{microtype}
\usepackage[hidelinks,unicode=true]{hyperref}
\hypersetup{colorlinks=true,% Don't draw a box around links, instead color every piece of text, which is a link
	linkcolor=blue,
	urlcolor=blue,
	citecolor=blue,
	pdfhighlight=/N% When clicking a link and holding the mouse button, don't use any graphical effects - i.e. the alternative would be to behave like a "button" which is pressed within the PDF
}%\usepackage{hyperref}

\newcommand{\av}[1]{\langle {#1} \rangle}

\usepackage[normalem]{ulem} 
\begin{document}

\begin{frontmatter}

\title{Adaptive network approach for emergence of societal bubbles}

\author[dpf]{H. P. Maia}
\author[dpf,nistcs]{S. C. Ferreira}
\author[dpf,nistcs,ibitphys]{M. L. Martins\corref{cor}}
\ead{mlmartins@ufv.br}

\address[dpf]{Departamento de F\'{\i}sica, Universidade Federal de Vi\c{c}osa, 36570-900, Vi\c{c}osa, MG, Brazil}
\address[nistcs]{National Institute of Science and technology for Complex Systems, Centro Brasileiro de Pesquisas F\'{\i}sicas, 
Rua Xavier Sigaud 150, 22290-180, Rio de Janeiro, Brazil}
\address[ibitphys]{Ibitipoca Institute of Physics - IbitiPhys, Concei\c{c}\~ao do Ibitipoca, 36140-000, MG, Brazil}

\cortext[cor]{Corresponding author.}

\date{\today}

\begin{abstract}
Far beyond its relevance for commercial and political marketings, opinion formation and
decision making processes are central for representative democracy, government
functioning, and state organization. In the present report, a stochastic agent-based model
is investigated. The model assumes that bounded confidence and homophily mechanisms drive
both opinion dynamics and social network evolution through either rewiring or breakage of
social contacts. In addition to the classical transition from global consensus to opinion
polarization, our main findings are (i) a cascade of fragmentation of the social network
into echo chambers (modules) holding distinct opinions and  rupture of the bridges
interconnecting these modules as the tolerance for opinion differences increases. There
are multiple surviving opinions associated to these modules within which consensus is
formed; and (ii) the adaptive social network exhibits a hysteresis{-like} behavior
characterized by irreversible changes in its topology as the opinion tolerance cycles from
radicalization towards  consensus  and backward to radicalization.
\end{abstract}


\end{frontmatter}

\section{Introduction}

Throughout human history, the government was ever exerted from a minority over
the majority and societies were organized through the game of oligarchies. After
the rise of capitalism, the socio-political stresses were mitigated by the
adoption and progressive enlargement of liberal democracy, an admirable system
that creates a majority and an opposition that agrees with the politics in
action~\cite{Hobsbawn}. Currently, the state power and power of wealth are 
engaged in reducing the political space~\cite{Ranciere}. Thus, dominant classes
controlling mass media and the information flow at the social networks, launched
out unprecedented efforts to limit the impact and to shape the opinion of the
majorities. The result, visible all around the world, is the the increasing of
political radicalization in a networked communicational environment, {catalyzing
	opinion polarization}. Media and social networks, mainly but not exclusively,
are forging pre-fabricated spaces for culture, religion, economy, law, and
policy~\cite{Deleuze}. In particular, mathematical models suggest that slightly
biased opinion surveys~\cite{Alves} and the action of agents with strategically
placed opinions~\cite{Hegselmann} can be determinant on the outcome of elections
or campaigns.

Here, we investigate opinion polarization dynamics from the viewpoint of complex
system theory. Essentially, we will be tracking the footprints of Dietrich
Stauffer, a pioneer in applications of statistical physics to socially motivated
problems \cite{Moss,Stauffer1}. In 2004, Stauffer and Meyer-Ortmanns
\cite{Stauffer2} {used extensive simulation analysis to obtain}  a critical 
value $\varepsilon_\text{c} \approx 0.4$ for the tolerance threshold (see
Sec.~\ref{sec:model}) for a complete consensus in the model of Deffuant
\textit{et al.} \cite{Deffuant}. In addition, they showed that the number of
different opinions when no complete consensus is formed is proportional to the
number of individuals. Since  pioneers {ignited the field}, the issue of
consensus versus polarization in opinion dynamics became a major topic in
sociophysics and many theoretical models were designed. See
Ref.~\cite{Castellano} for a seminal review of the related physics literature.
Examples include two of the most studied models in sociophysics, namely, the
bounded confidence model of Deffuant \textit{et al}.~\cite{Deffuant} and the
Sznajd model~\cite{Sznajd1}, the latter based on Ising spin variables. In
contrast to the opinion polarization observed in the Deffuant model and its
variants, the Sznajd model has only the complete consensus state as a dynamical
attractor in one- and two-dimensional regular lattices, as well as
complete graphs~\cite{Castellano}. Furthermore, the Sznajd model exhibits
bistability on the square lattice \cite{Sznajd2}: the initially majoritary
opinion by chance spreads throughout the whole population. However, despite
their different details, all these models reveal that even a small group of
extremists or zealots sustaining {long-term} fixed opinions may rule the whole
system dynamics and drive the majority opinion towards their own beliefs or
interests~\cite{Sznajd2,Mobilia}.

Concerning the nature of opinion radicalization and polarization phenomena,
Ramos~\textit{et al}.~\cite{Ramos} proposed a statistical
predictor of the rise of radicalization in society, which can be 
obtained from polls. The predictor is the onset of nonlinear behavior in the
scatter plots for the fractions of individuals holding a certain
extreme view versus the fraction of individuals sustaining either moderate or
that extreme opinions. A radicalization regime, characterized
by a wide spectrum of agents' opinions, emerges under large controversy and
strong social influence~\cite{Baumann}. Moreover, the introduction of homophily --- the
preference of individuals to interact with agents holding similar views ---
leads to a polarized state in which the opinion distribution function is
bimodal. It is worth to mention that emotions due to a conflict of opinions,
certainly enhanced by controversy, may harden individuals' and groups'
behaviors, generating a stable, conflicted social state~\cite{Sobkowicz1}.
Inflexible, extremists or zealot agents
{are not necessary} to trigger extremism rise. According to
	Sobkowicz~\cite{Sobkowicz2,Sobkowicz3}, ``Anyone can become an inflexible
zealot, anyone can become an extremist''. In turn, the role of media on public
opinion dynamics was considered in Ref.~\cite{Quattrociocchi}. It was shown
that, even at complete tolerance, plurality and competition within information
sources allows the stable coexistence of several and distinct cultures. In
contrast, an audience-oriented unpolarized media smoothes the transition from
polarization to consensus.

In all aforementioned works, the agents are nodes of homogeneous or
heterogeneous networks with fixed structures. So, the focus was ever on the
trajectories of the system states in well-defined phase spaces and opinion
dynamics does not affect the static network architecture. However, there are
many instances of networks whose states and topologies coevolve at similar time
scales. In these adaptive networks, state transitions and topological
alterations are inexorably intertwined, producing novel emergent
behaviors~\cite{Sayama}. In social phenomena, agents frequently modify their
contacts depending on the states assumed by the agents with which they interact.
In particular, person-to-person communication patterns and one-to many
information spreading channels, such as traditional media, blogs and microblogs
(like Twitter), evolve faster and faster. In the present work, we investigate a
modification of the  Deffuant model~\cite{Deffuant} on adaptive scale-free
networks in which network topology constrains agents' interactions and thus
opinion dynamics that, in turn, induces changes {in social interactions,
	altering the contact network}. Specifically, we quest for a fragmentation
transition analogous to that observed in the adaptive voter model
\cite{Kozma,Vazquez,Kimura}.

The paper is organized as follows. In Section~\ref{sec:model}, the stochastic
agent-based model is described. It is assumed that agent's opinions and the
social network structure coevolve in a subtle feedback loop regulated by bounded
confidence and homophily. In Section \ref{sec:results}, our major findings
obtained through extensive model simulations are reported. Particular emphasis
is given to the changes in topology of the adaptive social network driven by
opinion dynamics. These results are further discussed in
Section~\ref{sec:discussion}. Finally, some conclusions are drawn and
possible directions for future research are pointed out in
Section~\ref{sec:conclusions}.

\section{Model}
\label{sec:model}

The model is defined as follows. $N$ agents  are represented by the nodes of a
scale-free network. Each one has a {random} initial opinion $o_i(t=0)$,
{uniformly distributed in the interval $[0,1]$}. The  network of contacts at
$t=0$ is modeled using the uncorrelated configuration model
(UCM)~\cite{Catanzaro}, in which  each vertex is connected to $k_i(t=0)$ other
agents. The initial degrees $k_i(0)$, $i=1,2,\ldots,N$ are drawn at random
according a power-law degree distribution $P(k)\sim k^{-\gamma}$ with an upper
cutoff $k_\text{max}=\sqrt{N}$. Connections are performed at random  without
self nor multiple connections.  The UCM  generates synthetic,
undirected, and unweighted networks  without degree correlations for a tunable
degree exponent $\gamma$. In the simulations reported here, the value
$\gamma=2.7$ was used.

We consider a discrete time dynamics. At every time step $t>0$, agents' opinions
and their social contacts, thereby the network connectivity pattern,
coevolve under the following rules.

\textbf{Opinion dynamics}. Every agent $i$ compares his/her current opinion $o_i(t)$ with
the average opinion
\begin{equation}
\langle o \rangle_i(t)= \frac{1}{k_i(t)} \sum_{j \in \nu_i(t)} o_j(t)
\end{equation}
of his current social neighborhood $\nu_i(t)$, comprised by all the nodes $j$ connected to
$i$ at time $t$. Then, the difference $\Delta_i(t)=o_i(t)-\langle o \rangle_i(t)$ between
the agent's opinion and that of his social group is used to update the agent's view. The
rule is:

\begin{eqnarray} 
o_i(t+1)=\left\{
\begin{array}{lll}
o_i(t)+\mu \Delta_i(t) & \mbox{if }  &\Delta_i(t) \leq \varepsilon \\
o_i(t) & \mbox{if } & \Delta_i(t) > \varepsilon,\\
\end{array}
\right.
\label{eq:opinion_update}
\end{eqnarray}  
where  the parameter $\mu$ determines the speed of opinion convergence while
the parameter $\varepsilon$ is a tolerance threshold.  So, the opinion of the
agent $i$ tends to converge to that of his social group if they differ by less
than a value $\varepsilon$. This is the Deffuant et al.~\cite{Deffuant} formulation for
the bounded confidence principle: to be mutually influenced, agents (in our
version, an agent and his social group) must have similar enough opinions. If the
difference of opinions is too large, the communication process is impossible,
and opinions do not change. In the simulations, the value $\mu=0.8$ was fixed
while $\varepsilon$ was used as the control parameter of the model. All agent
opinions  are updated synchronously.

\begin{figure*}[htb]
	\centering
	\subfigure[\label{magnetization}]{\includegraphics[width=0.4\linewidth]{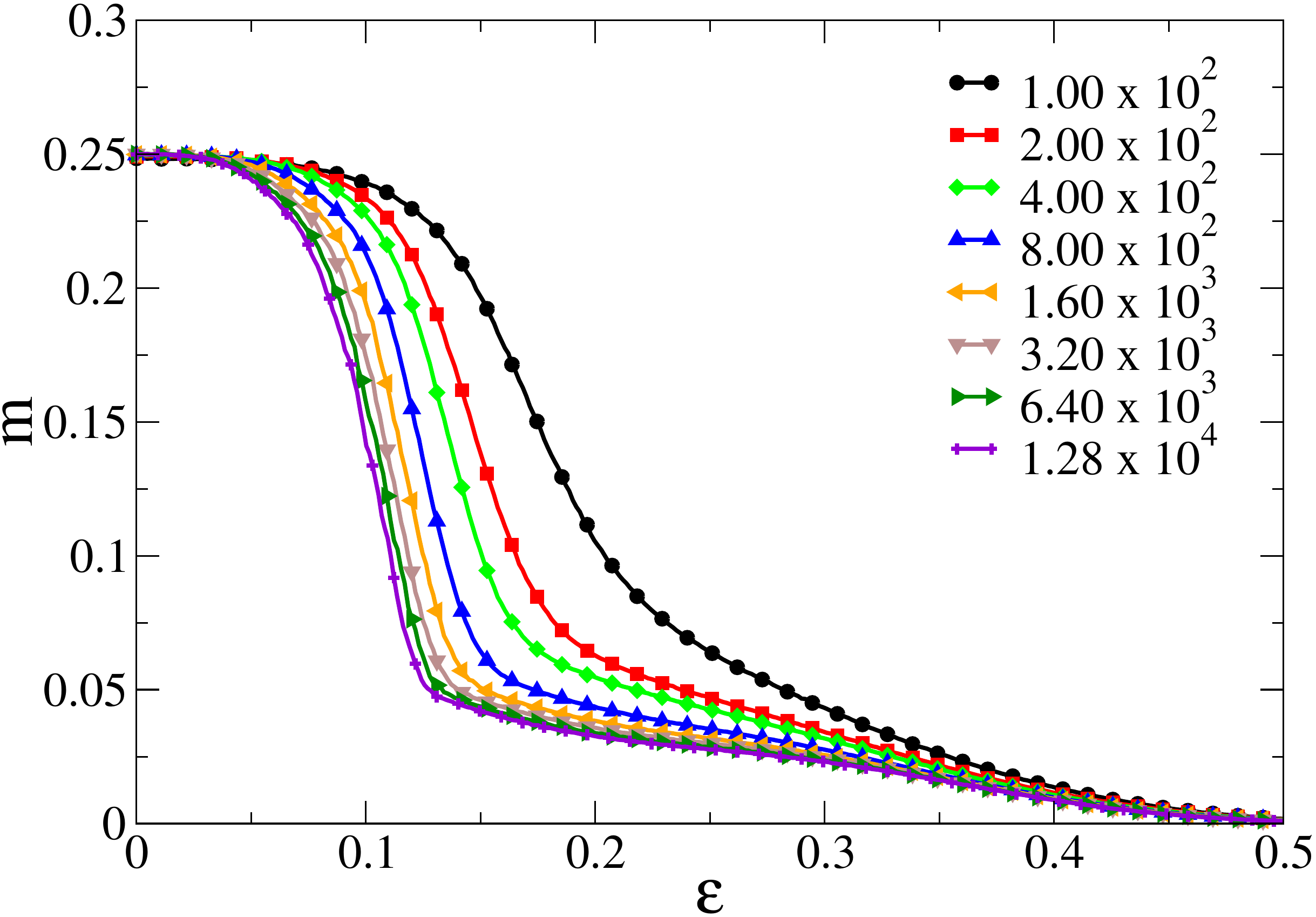}}
	\subfigure[\label{susceptibility}]{\includegraphics[width=0.4\linewidth]{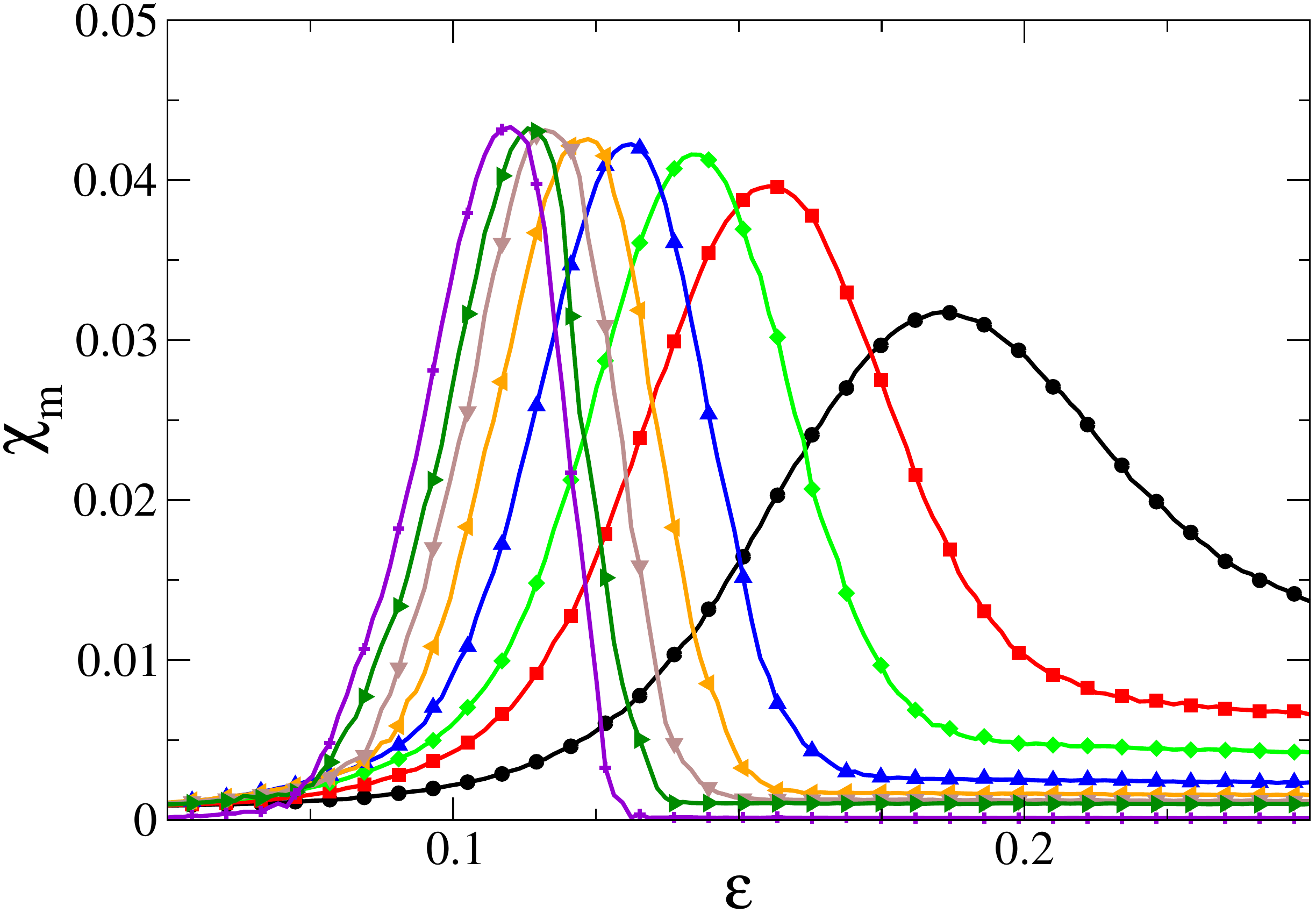}}
	\subfigure[\label{critical_epsilon}]{\includegraphics[width=0.4\linewidth]{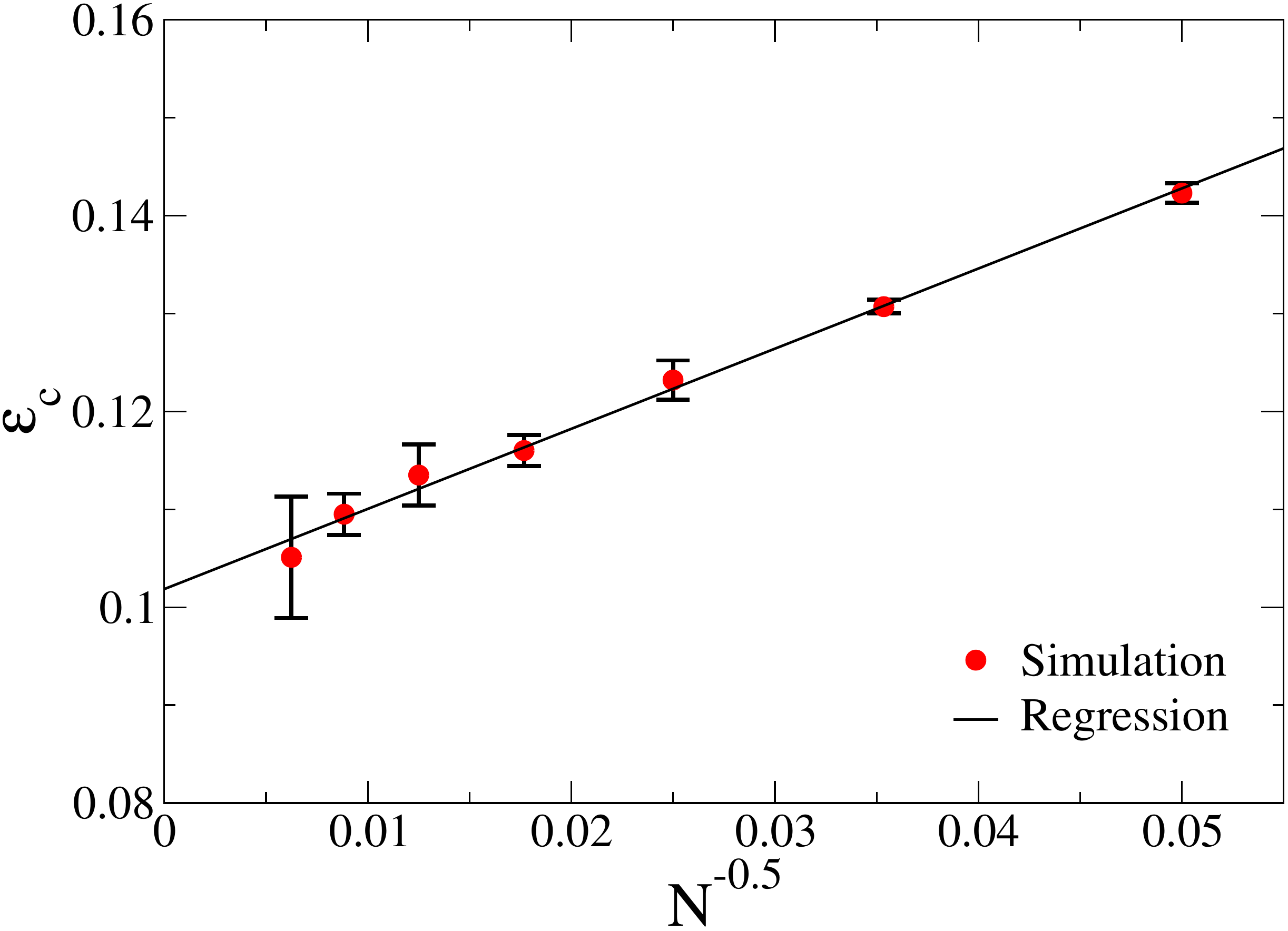}} 
	\caption{(a) Order parameter $m$ at the stationary state as a function of the tolerance
		threshold $\varepsilon$. (b) The stationary state opinions' variability
		$\chi_m=\av{o^2}_\text{en}-\av{o}_\text{en}^2$  computed over the ensembles of simulations for
		different system sizes $N$ indicated in the legend. (c) Finite-size scaling of
		$\varepsilon_\text{c}$, the critical tolerance threshold. Data correspond to averages over
		$500$ independent samples and time windows of $100$ steps after relaxation to the steady
		state.}
	\label{fig:order_parameter}
\end{figure*}

\textbf{Network topology dynamics}. After opinions' update, each edge
$e_{ij}(t)$ of the social network  can be broken and rewired (or not) according
to the following rules: (i) if the updated opinions of the agents connected by $e_{ij}$
differ in modulus by $\Delta_{ij}=|o_i(t+1)-o_j(t+1)|>\varepsilon$, the
corresponding edge $e_{ij}$ is broken with a probability $p=1-\exp(-\kappa
\Delta_{ij})$. The value $\kappa=3.5$ was fixed in the present simulations. In the
case of edge breakage, (ii) agents $i$ and $j$  rewire independently the broken
half-edge to other agents $l$ and $l^\prime$, respectively,  of similar opinions, i.
e., $\Delta_{il}, \Delta_{jl^\prime} \leq \varepsilon$. The agents $l$ and
$l^\prime$ are randomly chosen within a chemical distance $d_{ij}$ from $i$ and
$j$, and the edge rewiring probability $q$ decays with $d_{ij}$ as
\begin{eqnarray} 
q=\left\{
\begin{array}{ll}
\exp\left(-\frac{d_{ij}}{d_0}\right) &  \mbox{if }  d_{ij} \leq d_\text{max} \\
0 & \mbox{otherwise},\\
\end{array}
\right.
\label{eq:rewiring_probability}
\end{eqnarray}  
where $d_0$, a characteristic distance for rewiring, is a model parameter.  A
maximal distance for creation  of new contacts $d_\text{max} = \ln N$ was
adopted in the simulations. Its role is only to make the model
computationally more efficient avoiding the selection of nodes with $d_{ij}\gg
d_0$, for which the probability is essentially null.  Unless specified the distance
$d_0=4$ was used in all presented results . But, rewiring is ruled by
homophily, since agents have preference to interact with individuals holding
similar opinions. Finally, (iii)  not rewired half-edges can be replaced by new
contacts at later time steps according rule (ii). The limit $k_i(0)$, the
agent $i$ initial degree, is imposed only for the agent who searches a contact.

The rules forbid the  creation of links between agents whose opinions differ  above
the threshold tolerance $\varepsilon$. But in real societies very dissimilar
individuals still interact and is not rare to see extremists from the Left and
Right wings cooperating in politics, for example.  For this reason, we have also
tested a modified rule (ii): an agent can rewire a broken link to any individual
of the network with similar ($\Delta_{ij} < \varepsilon$) or dissimilar
($\Delta_{ij} > \varepsilon$) opinions   with fixed small probabilities $f$ and
$g$, respectively. This possibility 
allows for the interaction between individuals within
eventually distinct (isolated) clusters formed in the social network. Unless if
mentioned,  $f=0.01$ and $g=0.001$ were used for all simulations.

In summary, the present model for opinion dynamics involves two fundamental
processes  evolving at the same time scale: node states' changes as well as
breaking and rewiring of links. Therefore, this is an adaptive network
approach~\cite{Sayama,Castellano} where there is a coevolution of topology
adaptation and the opinion dynamics in the network, intertwined in a subtle
feedback loop.

\section{Results}
\label{sec:results}

Let us define a measure of consensus $m$ given by
\begin{equation}
m=\frac{1}{N} \sum_{i=1}^N |o_i(t)-\langle o \rangle(t)|,
\label{eq:order_parameter}
\end{equation} 
where $\langle o \rangle(t)=(1/N) \sum_i o_i(t)$ is the social average opinion
at time $t$. Figure \ref{fig:order_parameter}  presents the stationary behavior
of this quantity as a function of the tolerance threshold $\varepsilon$. As one
can see, $m$ behaves as an order parameter, vanishing above a critical threshold
$\varepsilon_\text{c}$. The variability associated to $m$ is defined as the
variance of $m$ computed over the ensemble of samples for $t\rightarrow\infty$,
$\chi_m= \av{o^2}_\text{en}-\av{o}_\text{en}^2$. This is measurement of the collective
opinions' fluctuations and shown in figure~\ref{susceptibility}. The curves $\chi_\text{m}$ versus $\varepsilon$
exhibit a peak  that can be used as an estimate of the transition point
$\varepsilon_\text{c}$ in analogy with a susceptibility which gives the
space-time fluctuations of a system~\cite{Marro2005}. At asymptotically large
$N$, a value $\varepsilon_\text{c} \approx 0.10$ was found using a finite size
scaling of the form
\begin{equation}
\varepsilon_\text{c}(N) = \varepsilon_\text{c}(\infty)+\text{const.}\times N^{-1/2},
\label{eq:fss}
\end{equation}
shown in figure~\ref{critical_epsilon}. 

\begin{figure}[hbt]
	\begin{center}
	\includegraphics[width=0.9\linewidth]{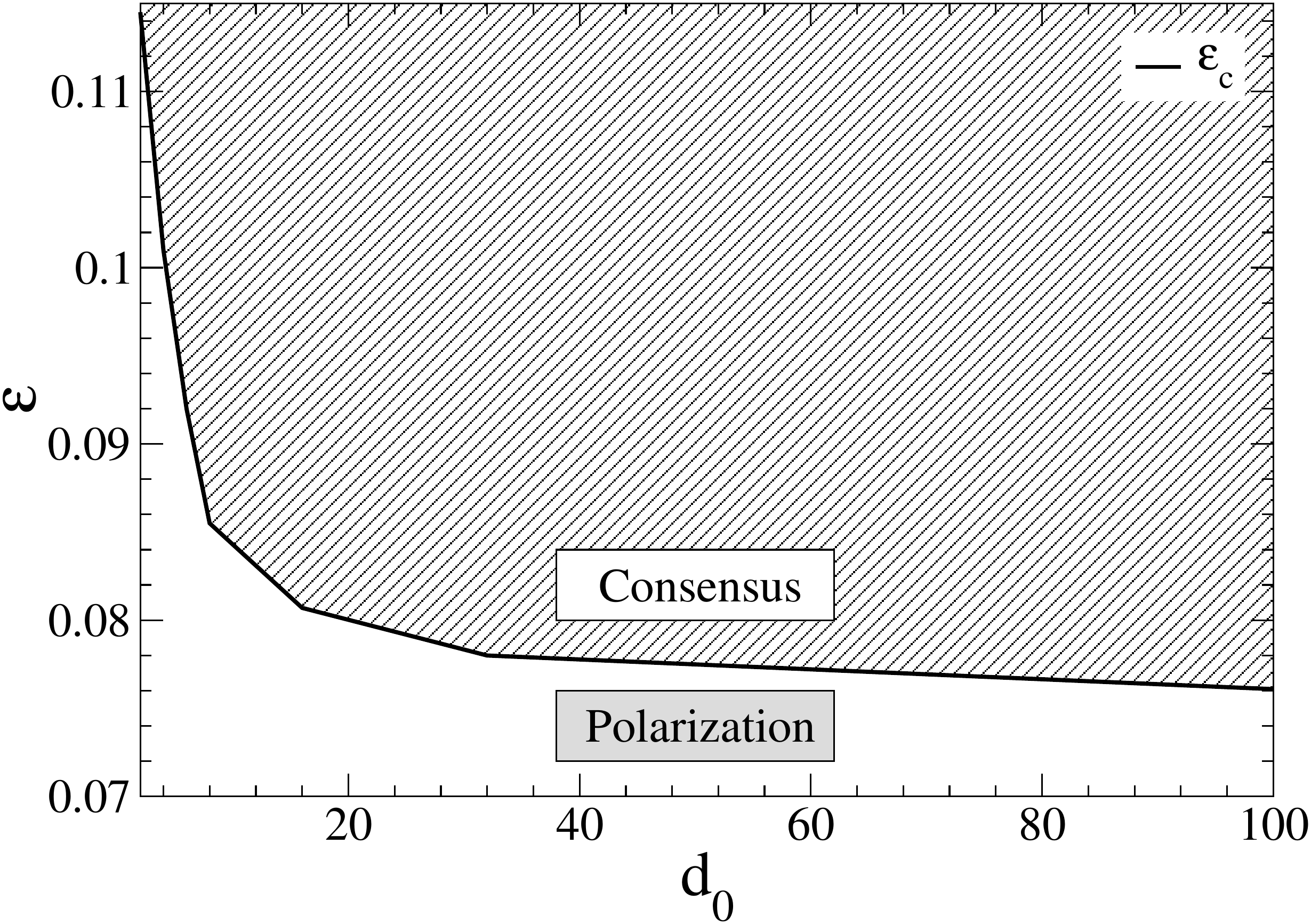}
	\end{center}
	\caption{Phase diagram in the parameter space $(\varepsilon,d_0)$ for the
		opinion dynamics model exhibiting a consensus-polarization transition. The separation line
		was determined using the finite-size scaling given by Eq.~\eqref{eq:fss}.  Averages were performed over $500$
		independent samples.}
	\label{phase_diagram}
\end{figure}

This transition corresponds to the known polarization-consensus transition
typical of models for opinion dynamics~\cite{Castellano}. Also, the critical
value $\varepsilon_\text{c} \approx 0.10$ we found is significantly lower than the values
$\varepsilon_\text{c} \approx 0.40$ and $\varepsilon_\text{c} \approx 0.50$ reported in
references~\cite{Stauffer2} and \cite{Fortunato} for static Barab\'asi-Albert
networks~\cite{Barabasi1999}, which are also scale-free with a degree exponent
$\gamma\approx 3$. Furthermore,  increasing characteristic distance $d_0$ for
rewiring of social contacts promotes full consensus formation by decreasing the
critical tolerance threshold (see Fig.~\ref{phase_diagram}). So, the transition between global
consensus and multipolarized (radicalized) states is mostly governed by the opinion threshold
tolerance $\varepsilon$ and the characteristic distance $d_0$ for social
contacts rewiring. A phase diagram in space parameter $\varepsilon$ versus
$d_0$, characterizing the polarized and consensus phases, is shown in
figure~\ref{phase_diagram}.

\begin{figure}[hbt]
\centering
\includegraphics[width=0.9\linewidth]{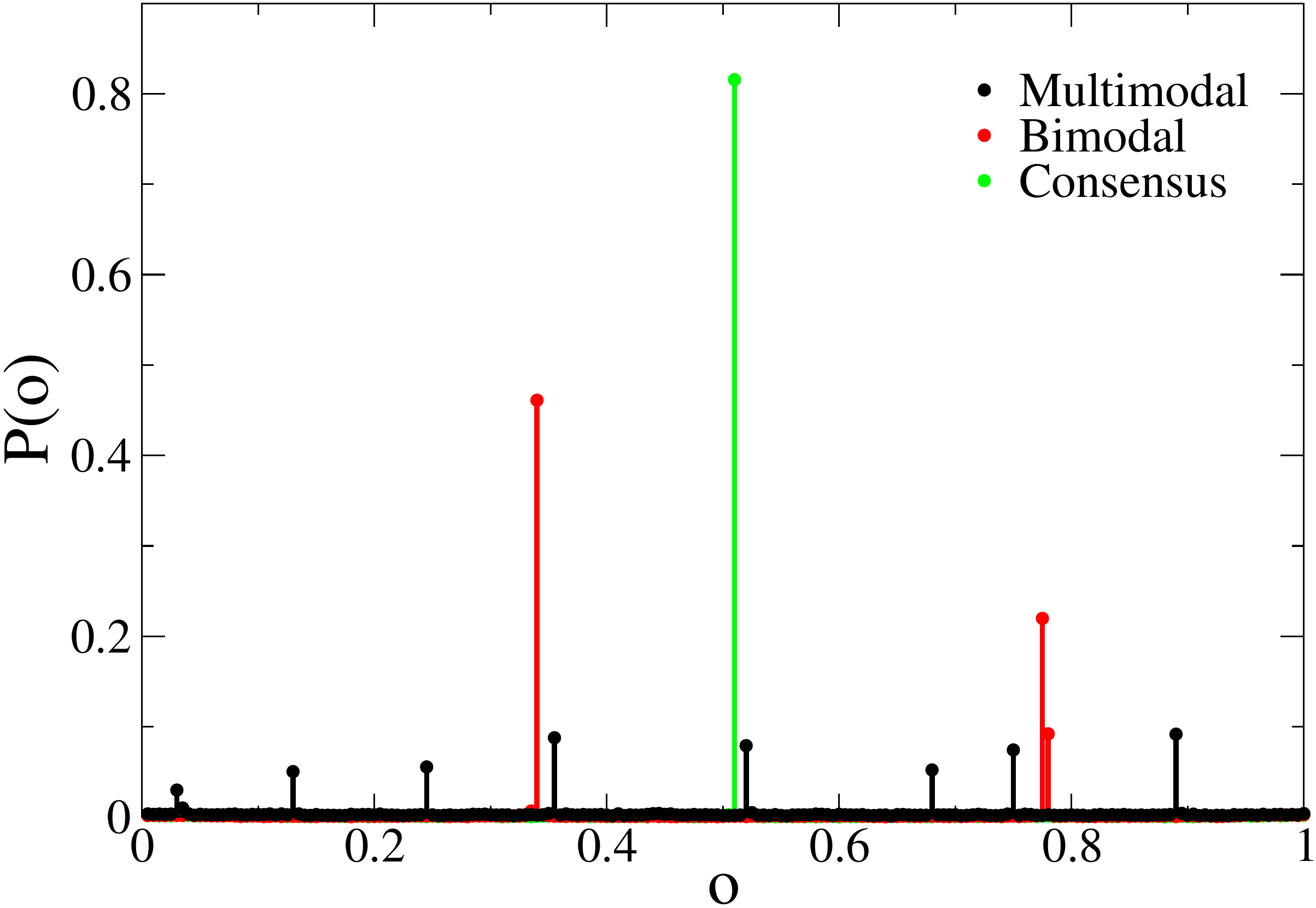}
\caption{Typical stationary distributions of the agents' opinion at the  full
	and majority consensus ($\varepsilon =0.11> \varepsilon_\text{c}$ - green), polarization (bimodal
	distribution - red), and radicalization (multimodal, even continuous distributions
	as $\varepsilon \rightarrow 0$ - black) regimes. The last two regimes occurs for
	$\varepsilon = 0.09$ and $\varepsilon=0.045$, respectively, both the critical tolerance threshold. }
	\label{OPD}
\end{figure}

In order to better illustrate the nature of the consensus-polarization
transition, the stationary opinion distributions for distinct tolerance
thresholds are shown in figure~\ref{OPD}. At the full consensus regime, observed
for large $\varepsilon$ values, this distribution is unimodal, since all the
agents share the same opinion. In contrast, at the polarization regime observed
for $\varepsilon < \varepsilon_\text{c}$, the consensus is destabilized, more opinions
emerge and their distribution is multimodal. Moreover, for $\varepsilon\ll
\varepsilon_\text{c}$, the number of surviving opinions at the stationary
state is proportional to the number of agents in the population, as communicated
by Stauffer and Meyer-Ortmanns~\cite{Stauffer2}.

\begin{figure*}[hbt]
	\subfigure[\label{fullconsensus_net}]{\includegraphics[width=0.24\linewidth]{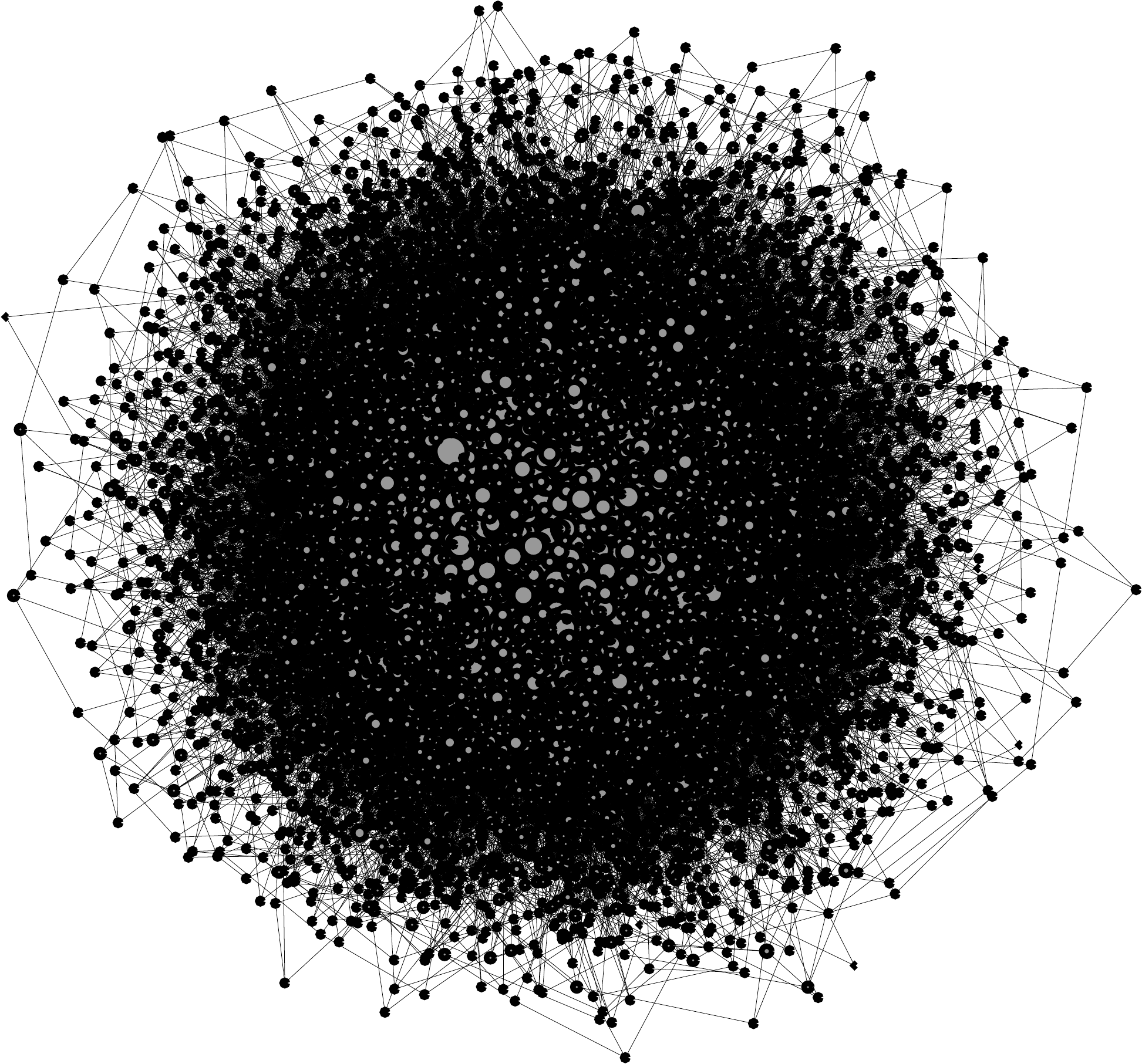}}
	\subfigure[\label{consensus_net}]{\includegraphics[width=0.24\linewidth]{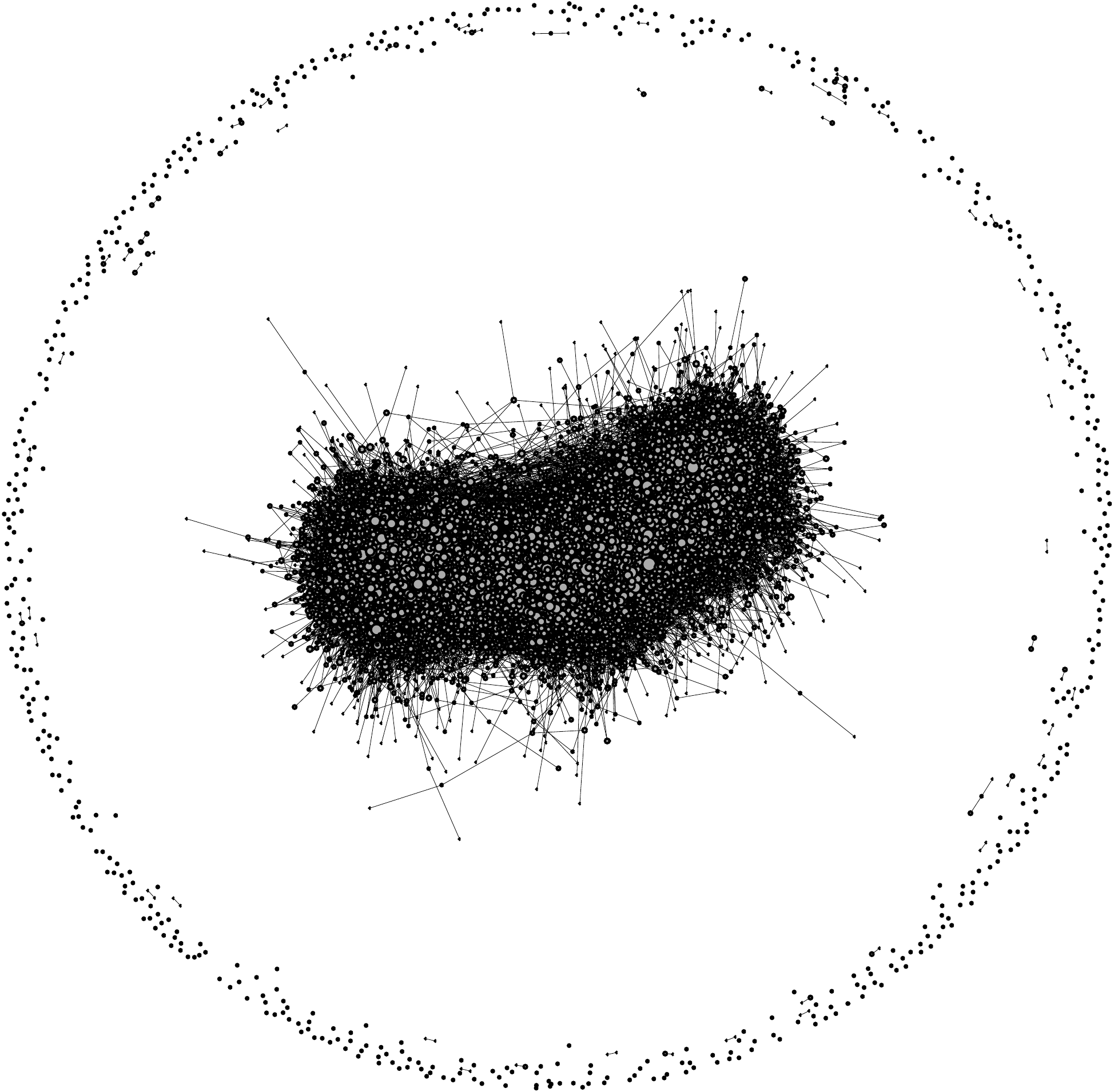}}
	\subfigure[\label{polarized_net}]{\includegraphics[width=0.24\linewidth]{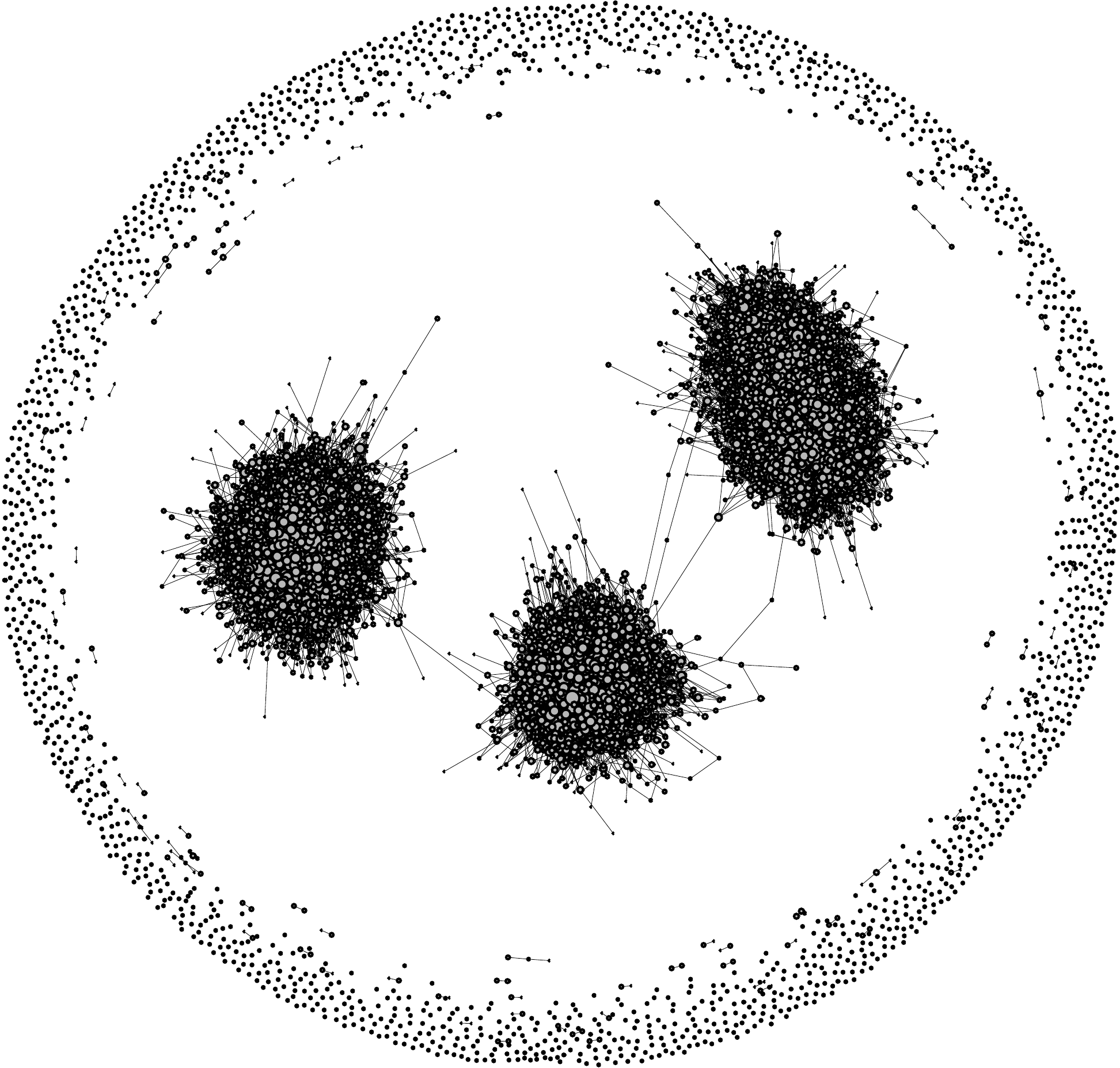}}
	\subfigure[\label{fragmented_net}]{\includegraphics[width=0.24\linewidth]{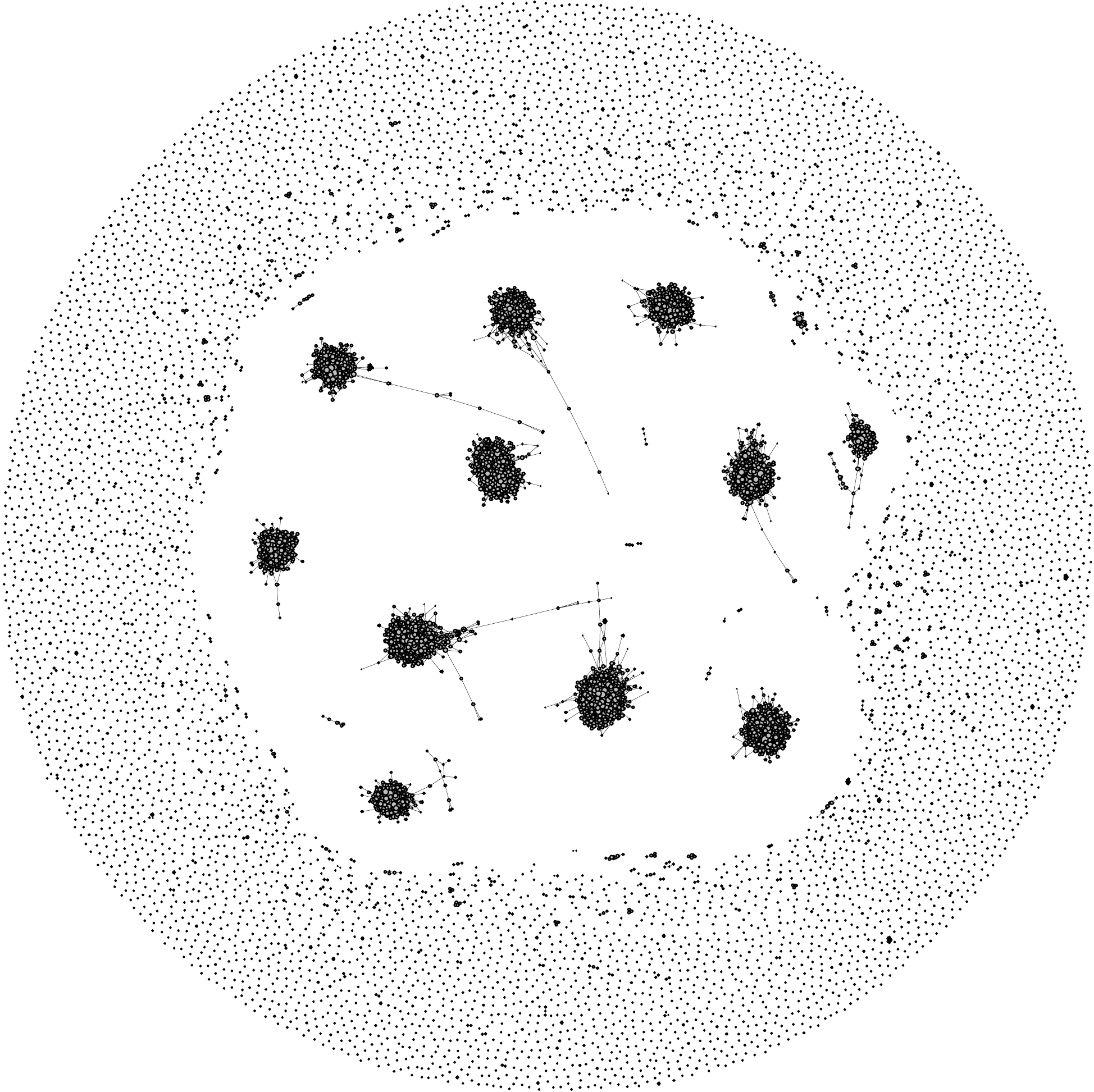}}
	\caption{Typical networks exhibiting (a) single community (the full consensus
		regime for $\varepsilon \ge 0.5$ in which $m=0$), (b) a single connected
		component concomitant with small isolated groups (majority consensus regime for
		$\varepsilon_c<\varepsilon<0.5$), (c) multiple modules connected by bridges (the
		radicalization regime for $\varepsilon \lesssim \varepsilon_\text{c}$), and (d)
		disconnected modules (still in the radicalization regime, but for  $\varepsilon
		\ll \varepsilon_{c}$). In this range, the rupture of several bridges between
		communities generates isolated clusters. These networks have $N=12,800$ nodes
		and tolerance thresholds are (a) $\varepsilon=0.5$, (b) $\varepsilon=0.145$, (c)
		$\varepsilon=0.085$, and (d) $\varepsilon=0.025$.}
	\label{fragmentation}
\end{figure*}

Concerning the structure (topology) of the adaptive social network, above
$\varepsilon_\text{c}$ the network is comprised by a {main, well connected} component and
several small components with a few nodes. In turn, the coevolution of opinions  and the
adaptive social network leads to the formation of communities within a modular structure
still in the non-fragmented network. Such modules were detected using the Louvain
algorithm~\cite{Blondel}. Slightly different leading opinions are initially present in each
community. Below $\varepsilon_\text{c}$, the polarization (bimodal opinion distribution)
or radicalization (multimodal opinion distribution) regimes emerge for
$\varepsilon\lesssim \varepsilon_\text{c}$ forming  still connected modular networks where
modules are interconnected by ``bridges''. As the tolerance threshold is   further reduced
the main component splits into two or more large components with sizes of the same order.
Within these isolated subgraphs, all agents share similar opinions but as $\varepsilon$ is
further reduced, these slightly different opinions lead to more module fragmentation. This
process continues until very low $\varepsilon$  when the networks is formed only by small
components. Such process is analogous to that observed in the adaptive voter
model~\cite{Kozma,Vazquez,Kimura}. A qualitative picture of this transition is shown in
figure~\ref{fragmentation}.

\begin{figure}[hbt]
	\begin{center}
		\includegraphics[width=0.9\linewidth]{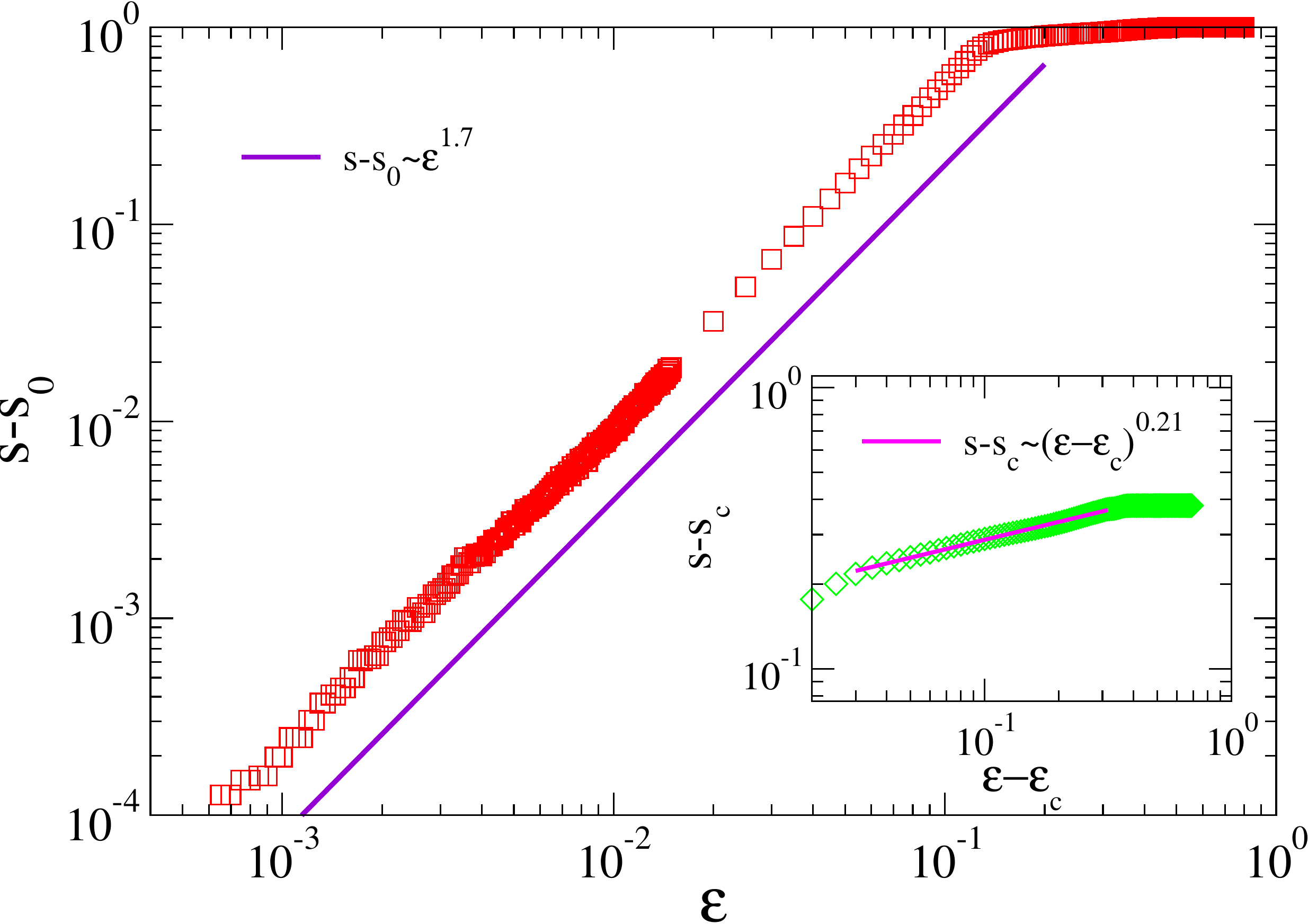}
	\end{center}
	\caption{Fractional size of the largest connected component as a function of the
		tolerance threshold $\varepsilon$. The constant $s_0=0.00135$ represents the average
		component fractional size as $\varepsilon\rightarrow 0$. Inset shows the scaling above the
		critical threshold. The network has $N=1,600$ nodes and averages were done over
		$10^4$ independent samples. }
	\label{S1}
\end{figure}

In addition to the number of surviving opinions in the polarization/radicalization
regimes, it is also relevant to know the size of the social groups holding those opinions.
Since opinion consensus (homogenization) involves agents in the same community or module,
we determined the size of the largest connected component $S_1$. Below
$\varepsilon_\text{c}$, the fraction $s=S_1/N$  scales very well as $ s -s_0 \sim
(\varepsilon_\text{c}-\varepsilon)^\delta$, where $s_0\approx 0.00135$  corresponds to the
fractional size of the largest component (of order $S=1$) as $\varepsilon\rightarrow 0$, and
$\delta\approx 1.7$; see figure~\ref{S1}. The approach to the critical tolerance threshold
for $\varepsilon>\varepsilon_\text{c}$  is also consistent  with a scaling law in the form
$s-s_c \sim (\varepsilon-\varepsilon_\text{c})^{\delta'}$ where $\delta'\approx 0.21$. 
Here, it worths to note that for $\epsilon>0.5$, we obtained $s= 1$  meaning that 
disconnected components are absent, in agreement with the transition found in
Ref.~\cite{Fortunato} for the Deffuant model.

\begin{figure*}[hbt]
	\centering
	\subfigure[\label{degrees}]{\includegraphics[width=0.4\linewidth]{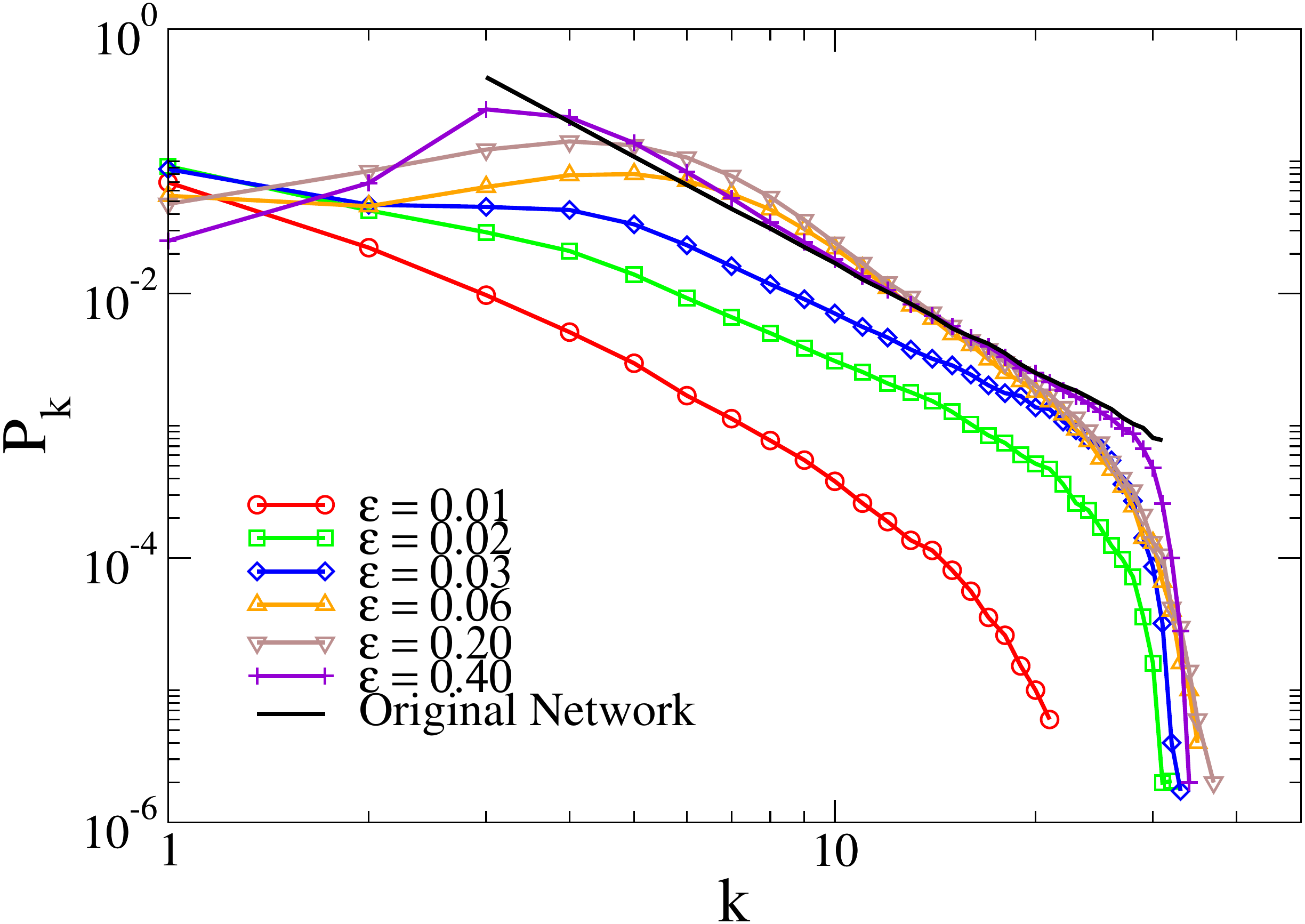}}
	\subfigure[\label{clustering}]{\includegraphics[width=0.4\linewidth]{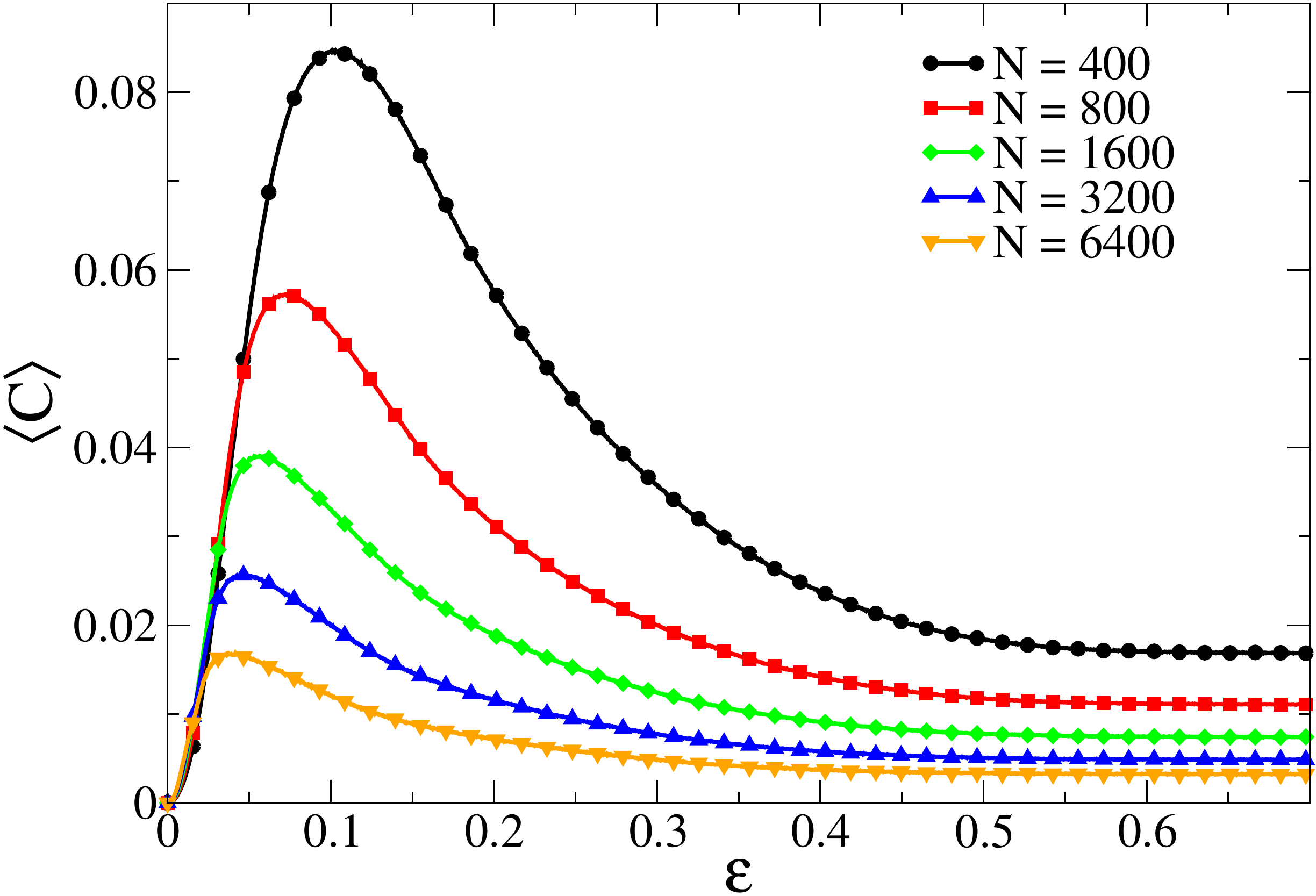}}\\
	\subfigure[\label{distance}]{\includegraphics[width=0.5\linewidth]{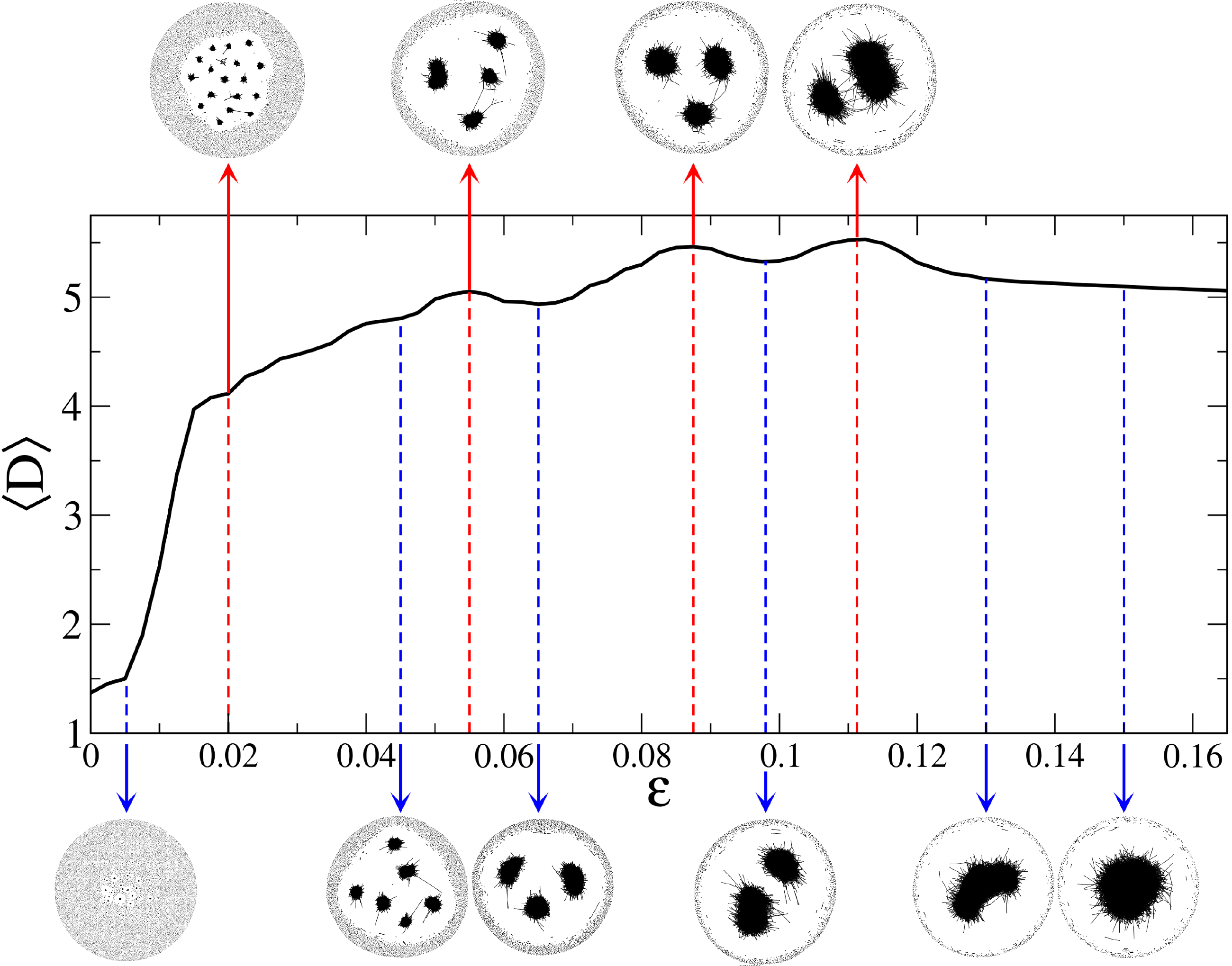}}
	\caption{(a) Typical degree distributions for stationary networks generated at distinct
		$\varepsilon$ values. $N=1000$ nodes was fixed.  (b) Clustering coefficient as a function
		of $\varepsilon$ for various network sizes $N$. Averages over $1000$ independent samples
		were performed. (c) Average distance between the nodes belonging to the largest connected
		component as a function of $\varepsilon$. The network size was fixed as $N=12,800$ nodes
		and averages were done over 1500 independent samples. Top and bottom snapshots represent
		typical network structures at minima and maxima of curve. }
	\label{topology_changes}
\end{figure*}

Both polarization and radicalization in opinion dynamics affect the adaptive network
topology through the breakage and rewiring of edges (social connections). A result is the
emergence of  a significant number of agents with small degrees and a few  nodes with
degrees larger than the UCM cut-off $k_\text{max}=N^{1/2}$~\cite{Catanzaro}. In
consequence, the network degree distribution deviates from the initial UCM power-law
degree distribution {only} at their tails (small and large degrees), except for
$\varepsilon\ll \varepsilon_\text{c}$ when the degree distribution becomes exponential in
the highly fragmented regime; see figure~\ref{degrees}. Due to homophily, individuals tend
to form ties with other individuals that also interact. This can be quantified by the
average clustering coefficient $\av{C}$~\cite{newman2010networks} of the network that
gives the average fraction of contacts of an individual which are themselves also
connected. Figure~\ref{clustering} presents the clustering coefficient found for contact
networks in the stationary state as function of the tolerance threshold. As expected,  the
initial null value of the clustering coefficient ($\av{C}\sim 1/N$) of the original UCM
network~\cite{Catanzaro} becomes finite when  the adaptive dynamics is at work. The
maximum value of $\av{C}$ occurs for $\varepsilon<\varepsilon_\text{c}$, whereas $\av{C}$
tends to zero in the limits $\varepsilon\rightarrow 0$ and $\varepsilon\gg
\varepsilon_\text{c}$. Changes of topology can be detected by the average shortest path
length $\av{D}$~\cite{newman2010networks} of the largest connected component as
illustrated in Fig.~\ref{distance}. The average distances present local maxima and minima
as $\varepsilon$ is increased from zero. The former occurs when the fragmentation of a
module into two large components is imminent, while the latter corresponds to the
emergence of groups with distinct opinions and community formation within a connected
fragment. Examples of network structures   for minima and maxima  shown in
figure~\ref{topology_changes} illustrate these structural changes.

\begin{figure}[ht]
	\begin{center}
		\includegraphics[width=0.99\linewidth]{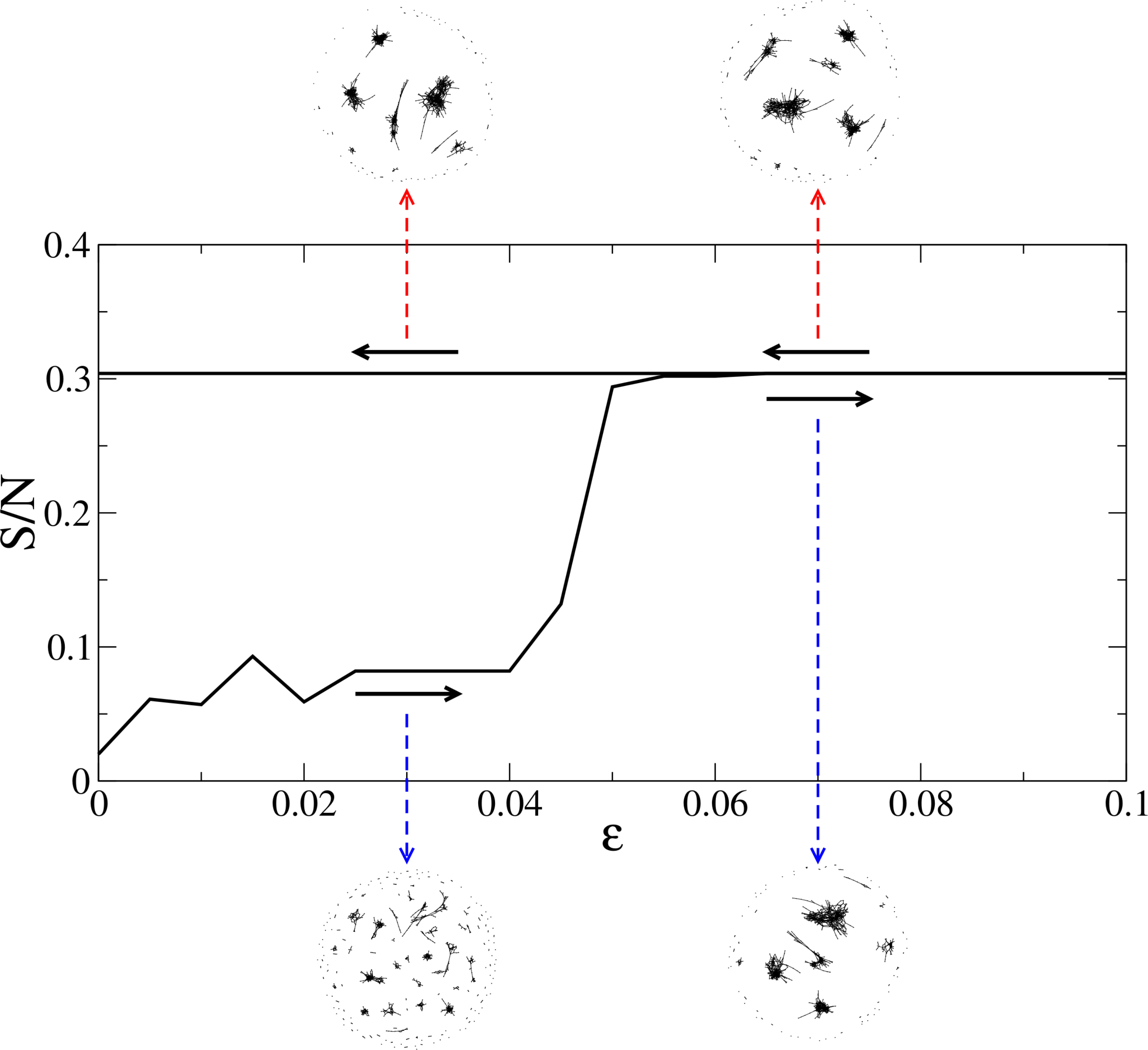}
	\end{center}
\caption{Hysteresis phenomena in the coevolution of opinions and the underlying social
	network structure. Typical stationary topologies are illustrated along the hysteresis
	loop. Network size  $N=10^3$ nodes, relaxation  $T=500$ times steps,  $f=0.2$, and
	$g=0.05$ were used.} 
\label{histeresis}
\end{figure}

The coevolution of opinions and the underlying adaptive social network exhibits a
hysteresis-like loop and, therefore, is irreversible below or across the polarization
transition. The $\varepsilon$-loop starts from an initial condition (opinions randomly
distributed on a UCM social network), which evolves up to its stationary state at fixed
$\varepsilon_0\approx 0$. Then, the stationary state (state variables and network
topology) becomes the initial state of the system, which now evolves at fixed
$\varepsilon_1=\varepsilon_0+\delta \varepsilon$ ($\delta \varepsilon \ll 1$).
This process is iterated $m$ times until $\varepsilon_m=\varepsilon_0+m \delta
\varepsilon$. So, the loop in $\varepsilon$ is closed decreasing $\varepsilon$ by $\delta
\varepsilon$ at each iteration, up to $\varepsilon_0$. As shown in figure
\ref{histeresis}, the network topology after the hysteresis loop is very different from
that at the starting of the loop since the system does not return to its initial
configuration.  In the first part of the $\varepsilon$-loop, modules or communities
interconnected by bridges become progressively more connected. On the way back,  once
consensus was formed, those connected components do not fragment as before, and we return
to a system (opinion distributions and network topology) certainly distinct from the
initial one. The impacts of such a phenomena in real societies can be huge.

\section{Discussion}
\label{sec:discussion}
Our coevolution model for the intertwined opinion and social network dynamics
revealed a very rich phenomenology. Indeed, four qualitatively dynamical
regimes, controlled mainly by the tolerance threshold $\varepsilon$, were found.
This parameter is crucial to generate a heterogeneous opinion distribution.
Starting from a random distribution of opinions, larger values yield full
consensus, whereas $\varepsilon \rightarrow 0$ yields an extreme heterogeneity
of opinions at the stationary state characterized by a uniform opinion
distribution function. The regimes and transitions/crossovers we have observed
exhibit the following traits.

\textbf{Full consensus}. At large tolerance threshold
values ($\varepsilon > 0.50$) a full consensus state, in which all agents hold
the same opinion is reached. In this regime, the social network is a graph
without isolated nodes or clusters and, therefore, the chemical distance
$d_{ij}$ between any pair $(i,j)$ of agents is finite. The network does
not have a modular structure and its topology (degree distribution, clustering
coefficient, average distance, etc.) is very similar to that of the initial UCM
network; see figure \ref{fullconsensus_net}. The onset of full consensus at
$\varepsilon \sim 0.50$ is consistent with the universal tolerance threshold for
complete consensus obtained in reference~\cite{Fortunato} for 
the Deffuant~\textit{ et al}. model in non-adaptive (fixed) networks.

\textbf{Majoritary consensus}. For intermediate values of the
tolerance threshold $0.10 < \varepsilon < 0.5$, an increasing, but small number
of agents disconnect  from the network. These agents stay isolated or eventually
form minute groups which  keep their opinion forever. So, the dynamics is ruled 
by a single and giant component of size hugely larger than that of all other
disconnected present on the network; see figure~\ref{consensus_net}. The
topology of the giant component is still similar to that of the initial UCM
network, but larger clustering coefficients are observed (see figure
\ref{clustering}) and the degree distribution deviates from the UCM counterpart
only at very small and large degree values. The full and majoritary consensus are
separated by a smooth crossover.

\noindent \textbf{Polarization}. For tolerance thresholds in the range $0.08
\lessapprox \varepsilon < 0.10$ the largest component becomes organized in
a modular structure containing two main communities. Further decreasing 
$\varepsilon$ makes these two modules progressively more evident and less
interconnected. Slightly below the tolerance threshold the
bridges linking  two loosely interconnected modules are broken (see figure
\ref{distance}). The emergence of a modular architecture renders the network
very distinct from the initial UCM graph. The  transition between majoritary
consensus and polarization at $\varepsilon_\text{c} \approx 0.10$ is characterized
by a non-divergent peak at the opinions' variability, figure \ref{susceptibility},
a local maximum in the average distance between nodes due to the split of the
population into two weakly connected modules, figure \ref{distance}, and a
crossover in the scaling behavior of the largest module size $s_1$; see
figure~\ref{S1}.

\noindent \textbf{Radicalization}. At the tolerance threshold $\varepsilon\approx
0.08$, the two large network components characterizing the polarization regime
fragment into three  weakly interconnected modules; see
figures~\ref{fragmented_net} and \ref{distance}. Further decrease of
$\varepsilon$ leads to a cascade of fragmentation and rupture of the bridges
interconnecting these modules. There are multiple surviving opinions associated
to these modules within which consensus is formed. The
polarization-radicalization crossover at is marked by a peak in the size of the
second largest component which is of same order as the largest one (data not
shown). In the range $0.008 < \varepsilon < 0.087$, there are several isolated
and loosely interconnected communities, significant in size, and  have large
clustering coefficients; see figure~\ref{clustering}. The independent dynamics
of these components is the cause of the polarization-radicalization transition.
Such a scenario is supported by references \cite{Cota2016,Cota2018} in which
extended control parameter regions with non-universal power-law decays of
activity in time were found in infinite dimensional, loosely coupled network of
modules. These ingredients indicate the existence of  Griffiths phases
\cite{Griffiths1969}. For $\varepsilon < 0.008$, the social network is pulverized
in a ``dust'' of isolated agents or very minute groups (the largest group size
is $S_1 \leq 5$ for $\varepsilon=0$), and the average distance between nodes
decays suddenly to $d \approx 1$ (see figure \ref{distance}). Also, the opinion
distribution function is almost continuous and its dynamics becomes effectively
frozen.

Once this general scenario was built, we will try to correlate it with recent phenomena
observed in real societies. On line communication networks, as Twitter and Facebook,
changed the way people behave, decide, opine and make choices~\cite{Masum}. These social
networks lead naturally to opinion polarization in communities with distinct views,
ultimately creating echo-chambers in which users reinforce their believes discussing with
each other within a closed, almost impermeable group. Social polarization and echo
chambers were detected in several contexts, namely, majoritary elections~\cite{Hanna2013},
political crisis~\cite{Wesley3}, street protests~\cite{Gonzalez,Borge}, and online
spreading of misinformation~\cite{Vicario}. This is exactly what happens in the
polarization and transition regimes of our model driven by homophily and operating at low
tolerance thresholds. In these regimes, the isolated or loosely interconnected modules
correspond to the echo-chambers, because within such modules consensus is reached and
their unanimous opinions can never meet each other. As in real societies, the surviving
opinions in the radicalization regime can not be rebutted in a democratic debate. The
challenge for liberal democracy is significant: A wrong politic can be widely accepted if
riding in one of the unanimity flows.

Furthermore, the hysteresis phenomenon reported here can have huge impacts on democracy.
Indeed, in democratic societies the government is under daily pressures imposed by the
multiplicity and heterogeneity of the organized interests of the social groups. These
demands are the  source of distinct political agendas, as well as ignite numerable and
simultaneous self-controlled combats. Growing radicalization (decreasing $\varepsilon$)
increases the social demands associated to the emergence of new group interests. This
means not only enhanced stresses on mass democracies, but also an increasing number of
potential discontented with the decisions and policies implemented. Such reverse effect of
democracy --- the number of upset interests tends to overcome the number of served
interests --- foster further radicalization and repudiation to the results of the
democratic practice. However, our results indicate that a reversal of radicalization trend
does not rescue the original, less fragmented (or radicalized) social network. 
According
our results, it is legitimate to speculate if the increase and differentiation of social
demands triggered by the outcomes of public policies and the reverse effects of democratic
competition are unavoidable in advanced capitalist societies grounded in universalized
political rights.

\section{Conclusions}
\label{sec:conclusions}
In this work we analyzed how opinion dynamics ruled by bounded confidence and
homophily can lead to social fragmentation and generate a society in bubbles. To
achieve this goal, extensive computer simulations of a stochastic agent-based
model, in which social contacts are broken and rewired with probabilities
dependent on the current opinion difference between two agents were performed.
The coevolution of opinion and social network topology driven by bounded
confidence and homophily revealed a very rich phenomenology characterized by
several dynamical regimes and continuous crossovers between them. We found that,
starting with a large tolerance threshold to opinion differences, the adaptive
social network progressively fragments in bubbles and the number of surviving
opinions increases as the tolerance threshold decreases. The bubbles are echo
chambers whose size distributions change from exponential with peaks at the tail
(large modules) to power-law and back to exponential, as the tolerance threshold
decreases. Within each bubble, consensus is formed. At the full consensus regime
(a unanimous opinion) the original network itself is the bubble. At the majoritary
consensus a single giant bubble rules the dynamics. This giant bubble organizes
into two modules that progressively disconnect from each other at the
polarization regime (two dominant opinions). At the radicalization regime
(multiple surviving opinions) the network is fragmented in several modules
(bubbles). We also found that the adaptive social network exhibits a
hysteresis-like behavior characterized by irreversible changes in its topology
as the opinion tolerance cycles from radicalization  consensus towards majoritary
and backward to radicalization consensus. The fragmentation in bubbles and its
associated irreversibility can have deep impact in majority formation and
democratic decision processes, as seems to be observed in contemporary real
societies. 

A forthcoming extension of the present model is to include the coevolution
dynamics of the tolerance threshold, the major control parameter of the model.
In addition, the effects of the rewiring characteristic distance and
probabilities ($d_0$, $f$, and $g$) calls for additional investigation. Finally,
extended versions of the present model, in which media and strategic agents are
included,  can be investigated to tackle the central issue of exogenous
influences on the opinion dynamics.

\section*{Acknowledgments}
Authors acknowledge the financial support of \textit{Conselho Nacional de Desenvolvimento
	Científico e Tecnológico} - CNPq (Grants no. 430768/2018-4 and 311183/2019-0) and
\textit{Fundação de Amparo à Pesquisa do Estado de Minas Gerais} - FAPEMIG  (Grant no.
APQ-02393-18).

%\bibliographystyle{elsarticle-num.bst}
%\bibliography{bib_society_in_bubbles}

\begin{thebibliography}{10}
	\expandafter\ifx\csname url\endcsname\relax
	\def\url#1{\texttt{#1}}\fi
	\expandafter\ifx\csname urlprefix\endcsname\relax\def\urlprefix{URL }\fi
	\expandafter\ifx\csname href\endcsname\relax
	\def\href#1#2{#2} \def\path#1{#1}\fi
	
	\bibitem{Hobsbawn}
	E.~J. Hobsbawn, The Age of Empire 1875-1914, Weidenfeld and Nicolson Ltd.,
	London, 1988.
	
	\bibitem{Ranciere}
	J.~Ranci\`ere, La haine de la d\'emocratie, La Fabrique \'Editions., Paris,
	2005.
	
	\bibitem{Deleuze}
	G.~Deleuze, Pourparlers, Les \'Editions de Minuit., Paris, 1990.
	
	\bibitem{Alves}
	S.~G. Alves, N.~M. {Oliveira Neto}, M.~L. Martins,
	\href{https://linkinghub.elsevier.com/retrieve/pii/S0378437102012086}{{Electoral
			surveys' influence on the voting processes: A cellular automata model}},
	Phys. A Stat. Mech. its Appl. 316~(1-4) (2002) 601--614.
	\newblock \href {http://dx.doi.org/10.1016/S0378-4371(02)01208-6}
	{\path{doi:10.1016/S0378-4371(02)01208-6}}.
	\newline\urlprefix\url{https://linkinghub.elsevier.com/retrieve/pii/S0378437102012086}
	
	\bibitem{Hegselmann}
	R.~Hegselmann, S.~K{\"{o}}nig, S.~Kurz, C.~Niemann, J.~Rambau,
	\href{http://jasss.soc.surrey.ac.uk/18/3/18.html}{{Optimal opinion control:
			The campaign problem}}, Jasss 18~(3) (2015) 18.
	\newblock \href {http://arxiv.org/abs/1410.8419} {\path{arXiv:1410.8419}},
	\href {http://dx.doi.org/10.18564/jasss.2847}
	{\path{doi:10.18564/jasss.2847}}.
	\newline\urlprefix\url{http://jasss.soc.surrey.ac.uk/18/3/18.html}
	
	\bibitem{Moss}
	S.~M. de~Oliveira, P.~M.~C. de~Oliveira, D.~Stauffer,
	\href{https://books.google.com.br/books?id=Ji-gBwAAQBAJ}{{Evolution, Money,
			War, and Computers: Non-Traditional Applications of Computational Statistical
			Physics}}, Teubner Texte zur Physik, Vieweg+Teubner Verlag, 2013.
	\newline\urlprefix\url{https://books.google.com.br/books?id=Ji-gBwAAQBAJ}
	
	\bibitem{Stauffer1}
	D.~Stauffer, \href{http://aip.scitation.org/doi/abs/10.1063/1.1632125}{{How to
			Convince Others? Monte Carlo Simulations of the Sznajd Model}}, AIP Conf.
	Proc. 690 (2003) 147--155.
	\newblock \href {http://arxiv.org/abs/0307133} {\path{arXiv:0307133}}, \href
	{http://dx.doi.org/10.1063/1.1632125} {\path{doi:10.1063/1.1632125}}.
	\newline\urlprefix\url{http://aip.scitation.org/doi/abs/10.1063/1.1632125}
	
	\bibitem{Stauffer2}
	D.~Stauffer, H.~Meyer-Ortmanns,
	\href{https://www.worldscientific.com/doi/abs/10.1142/S0129183104005644}{{Simulation
			of consensus model of deffuant et al. on a Barab{\'{a}}si- Albert network}},
	Int. J. Mod. Phys. C 15~(2) (2004) 241--246.
	\newblock \href {http://arxiv.org/abs/0308231} {\path{arXiv:0308231}}, \href
	{http://dx.doi.org/10.1142/S0129183104005644}
	{\path{doi:10.1142/S0129183104005644}}.
	\newline\urlprefix\url{https://www.worldscientific.com/doi/abs/10.1142/S0129183104005644}
	
	\bibitem{Deffuant}
	G.~Deffuant, D.~Neau, F.~Amblard, G.~Weisbuch,
	\href{https://www.worldscientific.com/doi/abs/10.1142/S0219525900000078}{{Mixing
			beliefs among interacting agents}}, Adv. Complex Syst. 03~(01n04) (2000)
	87--98.
	\newblock \href {http://dx.doi.org/10.1142/S0219525900000078}
	{\path{doi:10.1142/S0219525900000078}}.
	\newline\urlprefix\url{https://www.worldscientific.com/doi/abs/10.1142/S0219525900000078}
	
	\bibitem{Castellano}
	C.~Castellano, S.~Fortunato, V.~Loreto,
	\href{https://link.aps.org/doi/10.1103/RevModPhys.81.591}{{Statistical
			physics of social dynamics}}, Rev. Mod. Phys. 81~(2) (2009) 591--646.
	\newblock \href {http://arxiv.org/abs/0710.3256} {\path{arXiv:0710.3256}},
	\href {http://dx.doi.org/10.1103/RevModPhys.81.591}
	{\path{doi:10.1103/RevModPhys.81.591}}.
	\newline\urlprefix\url{https://link.aps.org/doi/10.1103/RevModPhys.81.591}
	
	\bibitem{Sznajd1}
	K.~Sznajd-Weron, J.~Sznajd,
	\href{https://www.worldscientific.com/doi/abs/10.1142/S0129183100000936}{{Opinion
			evolution in closed community}}, Int. J. Mod. Phys. C 11~(6) (2000)
	1157--1165.
	\newblock \href {http://arxiv.org/abs/0101130} {\path{arXiv:0101130}}, \href
	{http://dx.doi.org/10.1142/S0129183100000936}
	{\path{doi:10.1142/S0129183100000936}}.
	\newline\urlprefix\url{https://www.worldscientific.com/doi/abs/10.1142/S0129183100000936}
	
	\bibitem{Sznajd2}
	K.~Sznajd-Weron, M.~Tabiszewski, A.~M. Timpanaro,
	\href{https://iopscience.iop.org/article/10.1209/0295-5075/96/48002}{{Phase
			transition in the Sznajd model with independence}}, EPL (Europhysics Lett.
	96~(4) (2011) 48002.
	\newblock \href {http://arxiv.org/abs/1106.0934} {\path{arXiv:1106.0934}},
	\href {http://dx.doi.org/10.1209/0295-5075/96/48002}
	{\path{doi:10.1209/0295-5075/96/48002}}.
	\newline\urlprefix\url{https://iopscience.iop.org/article/10.1209/0295-5075/96/48002}
	
	\bibitem{Mobilia}
	M.~Mobilia, \href{https://link.aps.org/doi/10.1103/PhysRevLett.91.028701}{{Does
			a Single Zealot Affect an Infinite Group of Voters?}}, Phys. Rev. Lett.
	91~(2) (2003) 028701.
	\newblock \href {http://dx.doi.org/10.1103/PhysRevLett.91.028701}
	{\path{doi:10.1103/PhysRevLett.91.028701}}.
	\newline\urlprefix\url{https://link.aps.org/doi/10.1103/PhysRevLett.91.028701}
	
	\bibitem{Ramos}
	M.~Ramos, J.~Shao, S.~D.~S. Reis, C.~Anteneodo, J.~S. Andrade, S.~Havlin, H.~A.
	Makse, \href{http://www.nature.com/articles/srep10032}{{How does public
			opinion become extreme?}}, Sci. Rep. 5~(1) (2015) 10032.
	\newblock \href {http://arxiv.org/abs/1412.4718} {\path{arXiv:1412.4718}},
	\href {http://dx.doi.org/10.1038/srep10032} {\path{doi:10.1038/srep10032}}.
	\newline\urlprefix\url{http://www.nature.com/articles/srep10032}
	
	\bibitem{Baumann}
	F.~Baumann, P.~Lorenz-Spreen, I.~M. Sokolov, M.~Starnini,
	\href{https://link.aps.org/doi/10.1103/PhysRevLett.124.048301}{{Modeling Echo
			Chambers and Polarization Dynamics in Social Networks}}, Phys. Rev. Lett.
	124~(4) (2020) 048301.
	\newblock \href {http://arxiv.org/abs/1906.12325} {\path{arXiv:1906.12325}},
	\href {http://dx.doi.org/10.1103/PhysRevLett.124.048301}
	{\path{doi:10.1103/PhysRevLett.124.048301}}.
	\newline\urlprefix\url{https://link.aps.org/doi/10.1103/PhysRevLett.124.048301}
	
	\bibitem{Sobkowicz1}
	P.~Sobkowicz,
	\href{http://www.frontiersin.org/Interdisciplinary{\_}Physics/10.3389/fphy.2015.00017/abstract}{{Extremism
			without extremists: Deffuant model with emotions}}, Front. Phys. 3 (2015) 17.
	\newblock \href {http://dx.doi.org/10.3389/fphy.2015.00017}
	{\path{doi:10.3389/fphy.2015.00017}}.
	\newline\urlprefix\url{http://www.frontiersin.org/Interdisciplinary{\_}Physics/10.3389/fphy.2015.00017/abstract}
	
	\bibitem{Sobkowicz2}
	P.~Sobkowicz, \href{https://dx.plos.org/10.1371/journal.pone.0044489}{{Discrete
			Model of Opinion Changes Using Knowledge and Emotions as Control Variables}},
	PLoS One 7~(9) (2012) e44489.
	\newblock \href {http://dx.doi.org/10.1371/journal.pone.0044489}
	{\path{doi:10.1371/journal.pone.0044489}}.
	\newline\urlprefix\url{https://dx.plos.org/10.1371/journal.pone.0044489}
	
	\bibitem{Sobkowicz3}
	P.~Sobkowicz,
	\href{http://link.springer.com/10.1140/epjb/e2013-40029-0}{{Minority
			persistence in agent based model using information and emotional arousal as
			control variables}}, Eur. Phys. J. B 86~(7) (2013) 335.
	\newblock \href {http://dx.doi.org/10.1140/epjb/e2013-40029-0}
	{\path{doi:10.1140/epjb/e2013-40029-0}}.
	\newline\urlprefix\url{http://link.springer.com/10.1140/epjb/e2013-40029-0}
	
	\bibitem{Quattrociocchi}
	W.~Quattrociocchi, G.~Caldarelli, A.~Scala,
	\href{http://www.nature.com/articles/srep04938}{{Opinion dynamics on
			interacting networks: media competition and social influence}}, Sci. Rep.
	4~(1) (2015) 4938.
	\newblock \href {http://dx.doi.org/10.1038/srep04938}
	{\path{doi:10.1038/srep04938}}.
	\newline\urlprefix\url{http://www.nature.com/articles/srep04938}
	
	\bibitem{Sayama}
	H.~Sayama, I.~Pestov, J.~Schmidt, B.~J. Bush, C.~Wong, J.~Yamanoi, T.~Gross,
	\href{https://linkinghub.elsevier.com/retrieve/pii/S0898122112007018}{{Modeling
			complex systems with adaptive networks}}, Comput. Math. with Appl. 65~(10)
	(2013) 1645--1664.
	\newblock \href {http://dx.doi.org/10.1016/j.camwa.2012.12.005}
	{\path{doi:10.1016/j.camwa.2012.12.005}}.
	\newline\urlprefix\url{https://linkinghub.elsevier.com/retrieve/pii/S0898122112007018}
	
	\bibitem{Kozma}
	B.~Kozma, A.~Barrat,
	\href{https://link.aps.org/doi/10.1103/PhysRevE.77.016102}{{Consensus
			formation on adaptive networks}}, Phys. Rev. E 77~(1) (2008) 016102.
	\newblock \href {http://arxiv.org/abs/0707.4416} {\path{arXiv:0707.4416}},
	\href {http://dx.doi.org/10.1103/PhysRevE.77.016102}
	{\path{doi:10.1103/PhysRevE.77.016102}}.
	\newline\urlprefix\url{https://link.aps.org/doi/10.1103/PhysRevE.77.016102}
	
	\bibitem{Vazquez}
	F.~Vazquez, V.~M. Egu{\'{i}}luz, M.~S. Miguel,
	\href{https://link.aps.org/doi/10.1103/PhysRevLett.100.108702}{{Generic
			Absorbing Transition in Coevolution Dynamics}}, Phys. Rev. Lett. 100~(10)
	(2008) 108702.
	\newblock \href {http://arxiv.org/abs/0710.4910} {\path{arXiv:0710.4910}},
	\href {http://dx.doi.org/10.1103/PhysRevLett.100.108702}
	{\path{doi:10.1103/PhysRevLett.100.108702}}.
	\newline\urlprefix\url{https://link.aps.org/doi/10.1103/PhysRevLett.100.108702}
	
	\bibitem{Kimura}
	D.~Kimura, Y.~Hayakawa,
	\href{https://link.aps.org/doi/10.1103/PhysRevE.78.016103}{{Coevolutionary
			networks with homophily and heterophily}}, Phys. Rev. E 78~(1) (2008) 016103.
	\newblock \href {http://dx.doi.org/10.1103/PhysRevE.78.016103}
	{\path{doi:10.1103/PhysRevE.78.016103}}.
	\newline\urlprefix\url{https://link.aps.org/doi/10.1103/PhysRevE.78.016103}
	
	\bibitem{Catanzaro}
	M.~Catanzaro, M.~Bogu{\~{n}}{\'{a}}, R.~Pastor-Satorras,
	\href{https://link.aps.org/doi/10.1103/PhysRevE.71.027103}{{Generation of
			uncorrelated random scale-free networks}}, Phys. Rev. E 71 (2005) 027103.
	\newblock \href {http://dx.doi.org/10.1103/PhysRevE.71.027103}
	{\path{doi:10.1103/PhysRevE.71.027103}}.
	\newline\urlprefix\url{https://link.aps.org/doi/10.1103/PhysRevE.71.027103}
	
	\bibitem{Marro2005}
	J.~Marro, R.~Dickman,
	\href{https://books.google.com.br/books?id=80YF69jbczYC}{{Nonequilibrium
			Phase Transitions in Lattice Models}}, Al{\'{e}}a-Saclay, Cambridge
	University Press, 2005.
	\newline\urlprefix\url{https://books.google.com.br/books?id=80YF69jbczYC}
	
	\bibitem{Fortunato}
	S.~Fortunato,
	\href{https://www.worldscientific.com/doi/abs/10.1142/S0129183104006728}{{Universality
			of the threshold for complete consensus for the opinion dynamics of Deffuant
			et al.}}, Int. J. Mod. Phys. C 15 (2004) 1301--1307.
	\newblock \href {http://dx.doi.org/10.1142/S0129183104006728}
	{\path{doi:10.1142/S0129183104006728}}.
	\newline\urlprefix\url{https://www.worldscientific.com/doi/abs/10.1142/S0129183104006728}
	
	\bibitem{Barabasi1999}
	A.~L. Barab{\'{a}}si, R.~Albert,
	\href{http://www.sciencemag.org/cgi/doi/10.1126/science.286.5439.509}{{Emergence
			of scaling in random networks}}, Science 286~(5439) (1999) 509--512.
	\newblock \href {http://dx.doi.org/10.1126/science.286.5439.509}
	{\path{doi:10.1126/science.286.5439.509}}.
	\newline\urlprefix\url{http://www.sciencemag.org/cgi/doi/10.1126/science.286.5439.509}
	
	\bibitem{Blondel}
	V.~D. Blondel, J.-L. Guillaume, R.~Lambiotte, E.~Lefebvre,
	\href{https://iopscience.iop.org/article/10.1088/1742-5468/2008/10/P10008}{{Fast
			unfolding of communities in large networks}}, J. Stat. Mech. Theory Exp.
	2008~(10) (2008) P10008.
	\newblock \href {http://arxiv.org/abs/0803.0476} {\path{arXiv:0803.0476}},
	\href {http://dx.doi.org/10.1088/1742-5468/2008/10/P10008}
	{\path{doi:10.1088/1742-5468/2008/10/P10008}}.
	\newline\urlprefix\url{https://iopscience.iop.org/article/10.1088/1742-5468/2008/10/P10008}
	
	\bibitem{newman2010networks}
	M.~Newman,
	\href{http://www.oxfordscholarship.com/view/10.1093/acprof:oso/9780199206650.001.0001/acprof-9780199206650}{{Networks}},
	Oxford University Press, Oxford New York, 2010.
	\newblock \href {http://dx.doi.org/10.1093/acprof:oso/9780199206650.001.0001}
	{\path{doi:10.1093/acprof:oso/9780199206650.001.0001}}.
	\newline\urlprefix\url{http://www.oxfordscholarship.com/view/10.1093/acprof:oso/9780199206650.001.0001/acprof-9780199206650}
	
	\bibitem{Cota2016}
	W.~Cota, S.~C. Ferreira, G.~{\'{O}}dor,
	\href{http://link.aps.org/doi/10.1103/PhysRevE.93.032322}{{Griffiths effects
			of the susceptible-infected-susceptible epidemic model on random power-law
			networks}}, Phys. Rev. E 93 (2016) 032322.
	\newblock \href {http://dx.doi.org/10.1103/PhysRevE.93.032322}
	{\path{doi:10.1103/PhysRevE.93.032322}}.
	\newline\urlprefix\url{http://link.aps.org/doi/10.1103/PhysRevE.93.032322}
	
	\bibitem{Cota2018}
	W.~Cota, G.~{\'{O}}dor, S.~C. Ferreira,
	\href{http://www.nature.com/articles/s41598-018-27506-x}{{Griffiths phases in
			infinite-dimensional, non-hierarchical modular networks}}, Sci. Rep. 8~(1)
	(2018) 9144.
	\newblock \href {http://dx.doi.org/10.1038/s41598-018-27506-x}
	{\path{doi:10.1038/s41598-018-27506-x}}.
	\newline\urlprefix\url{http://www.nature.com/articles/s41598-018-27506-x}
	
	\bibitem{Griffiths1969}
	R.~B. Griffiths,
	\href{http://link.aps.org/doi/10.1103/PhysRevLett.23.17}{{Nonanalytic
			Behavior Above the Critical Point in a Random Ising Ferromagnet}}, Phys. Rev.
	Lett. 23~(1) (1969) 17--19.
	\newblock \href {http://dx.doi.org/10.1103/PhysRevLett.23.17}
	{\path{doi:10.1103/PhysRevLett.23.17}}.
	\newline\urlprefix\url{http://link.aps.org/doi/10.1103/PhysRevLett.23.17}
	
	\bibitem{Masum}
	H.~Masum, M.~Tovey, C.~Newmark,
	\href{https://books.google.com.br/books?id=Mk81XgHfh1cC}{{The Reputation
			Society: How Online Opinions Are Reshaping the Offline World}}, The
	Information Society Series, MIT Press, 2012.
	\newline\urlprefix\url{https://books.google.com.br/books?id=Mk81XgHfh1cC}
	
	\bibitem{Hanna2013}
	A.~Hanna, C.~Wells, P.~Maurer, L.~Friedland, D.~Shah, J.~Matthes,
	\href{http://dl.acm.org/citation.cfm?doid=2508436.2508438}{{Partisan
			alignments and political polarization online}}, in: Proc. 2nd Work. Polit.
	elections data - PLEAD '13, ACM Press, New York, New York, USA, 2013, pp.
	15--22.
	\newblock \href {http://dx.doi.org/10.1145/2508436.2508438}
	{\path{doi:10.1145/2508436.2508438}}.
	\newline\urlprefix\url{http://dl.acm.org/citation.cfm?doid=2508436.2508438}
	
	\bibitem{Wesley3}
	W.~Cota, S.~C. Ferreira, R.~Pastor-Satorras, M.~Starnini,
	\href{https://epjdatascience.springeropen.com/articles/10.1140/epjds/s13688-019-0213-9}{{Quantifying
			echo chamber effects in information spreading over political communication
			networks}}, EPJ Data Sci. 8~(1) (2019) 35.
	\newblock \href {http://arxiv.org/abs/1901.03688} {\path{arXiv:1901.03688}},
	\href {http://dx.doi.org/10.1140/epjds/s13688-019-0213-9}
	{\path{doi:10.1140/epjds/s13688-019-0213-9}}.
	\newline\urlprefix\url{https://epjdatascience.springeropen.com/articles/10.1140/epjds/s13688-019-0213-9}
	
	\bibitem{Gonzalez}
	S.~Gonz{\'{a}}lez-Bail{\'{o}}n, J.~Borge-Holthoefer, A.~Rivero, Y.~Moreno,
	\href{http://www.nature.com/articles/srep00197}{{The Dynamics of Protest
			Recruitment through an Online Network}}, Sci. Rep. 1~(1) (2011) 197.
	\newblock \href {http://dx.doi.org/10.1038/srep00197}
	{\path{doi:10.1038/srep00197}}.
	\newline\urlprefix\url{http://www.nature.com/articles/srep00197}
	
	\bibitem{Borge}
	J.~Borge-Holthoefer, W.~Magdy, K.~Darwish, I.~Weber,
	\href{http://dl.acm.org/citation.cfm?doid=2675133.2675163}{{Content and
			Network Dynamics Behind Egyptian Political Polarization on Twitter}}, Proc.
	18th ACM Conf. Comput. Support. Coop. Work Soc. Comput. - CSCW '15 (2015)
	700--711\href {http://arxiv.org/abs/1410.3097} {\path{arXiv:1410.3097}},
	\href {http://dx.doi.org/10.1145/2675133.2675163}
	{\path{doi:10.1145/2675133.2675163}}.
	\newline\urlprefix\url{http://dl.acm.org/citation.cfm?doid=2675133.2675163}
	
	\bibitem{Vicario}
	M.~{Del Vicario}, G.~Vivaldo, A.~Bessi, F.~Zollo, A.~Scala, G.~Caldarelli,
	W.~Quattrociocchi, \href{http://www.nature.com/articles/srep37825}{{Echo
			Chambers: Emotional Contagion and Group Polarization on Facebook}}, Sci. Rep.
	6~(1) (2016) 37825.
	\newblock \href {http://arxiv.org/abs/1607.01032} {\path{arXiv:1607.01032}},
	\href {http://dx.doi.org/10.1038/srep37825} {\path{doi:10.1038/srep37825}}.
	\newline\urlprefix\url{http://www.nature.com/articles/srep37825}
	
\end{thebibliography}

\end{document}